\documentstyle[11pt,epsfig]{article}
\newcommand{\Bbb}{\cal}
\newcommand{\psbild}[5]
{\par
 \begin{figure}[#1]
 \begin{center}
 \begin{minipage}{0cm} \end{minipage}
 \begin{minipage}{#3}
 \refstepcounter{figure}\label{#2}
 \epsfxsize=#3
 \epsffile{#4}
 \end{minipage}
 \end{center}
 \hfill
 \begin{minipage}{0cm} \end{minipage}
 \begin{center}
 \parbox{15cm}{\baselineskip11pt{\rm Figure {#2}}: {#5}}
 \end{center}
 \end{figure}} 
\newcommand{\resection}[1]{\setcounter{equation}{0}\section{#1}}

\def\Trace{{\rm Trace\,}}
\def\Im{{\rm Im\,}}
\def\Re{{\rm Re\,}}
\newcommand{\beq}{\begin{equation}}
\newcommand{\eeq}{\end{equation}}
\newcommand{\bdm}{\begin{displaymath}}
\newcommand{\edm}{\end{displaymath}}
\newcommand{\bea}{\begin{eqnarray}}
\newcommand{\eea}{\end{eqnarray}}

\begin{document}
\setcounter{page}{0}
\topmargin 0pt
\newpage
\setcounter{page}{0}
\begin{titlepage}
\vspace{0.5cm}
\begin{flushright}{SWAT 95-96/115}\\
Maths 96/2
\end{flushright}
\begin{center}
{\Large 
{\bf Inverse scattering and solitons in $A_{n-1}$ affine Toda field theories}} \\
\vspace{1cm}
{\large Edwin J. Beggs$^*$ and Peter R. Johnson$^\dagger$}\\
\vspace{1.0cm}
{\em $^*$Department of Mathematics,\\
University of Wales at Swansea,\\
Singleton Park,\\
Swansea,\\
 SA2 8PP, Wales, UK.} \\ 
\vspace{0.2 cm}
{\em $^\dagger$Department of Physics,\\
University of Wales at Swansea,\\
Singleton Park,\\
Swansea,\\
 SA2 8PP, Wales, UK.} \\ \end{center}
\vspace{1cm}
\vspace{0.5cm}
\setcounter{footnote}{0}\baselineskip=14pt 
\begin{center}{\bf Abstract} \\ \end{center}

We implement the inverse scattering method in the case of 
the $A_n$ affine Toda
field theories, by studying the space-time evolution of simple poles
in the underlying loop group. We find the known single and
multi-soliton solutions, as well as additional solutions with
non-linear modes of oscillation around the standard solution, by
studying the particularly simple case where the residue at the pole is a rank
one projection. We show that these solutions with extra modes
have the same mass and topological charges as the standard solutions,
so we do not shed any light on the missing topological charge problem
in these models. We also show that the integrated energy-momentum
density can be calculated from the central extension of the loop group.
\end{titlepage}\baselineskip=14pt 
\resection{Introduction}
This paper applies the inverse scattering method to the $A_n$ affine
Toda field theories. The inverse scattering method employs complex analysis
to solve integrable models, for example the principal chiral model 
\cite{ZS1,ZS2,Novikov,Edwin} and the sine-Gordon model
\cite{Novikov,Faddeev}. The $su(n)$ Toda lattice was considered
in \cite{Mik}. 
The method depends on the existence of a linear system, which contains 
a complex parameter $\lambda$, called the spectral parameter. This can
be thought of as a co-ordinate around a circle (when $\lambda\in 
{\Bbb R}\cup\{\infty\}$), giving rise to a loop group \cite{PS}.

In this picture, the solitons  are generated by elements of the loop
group which are meromorphic functions of the spectral parameter.
Multiplication by a meromorphic loop introduces extra solitons into 
the system. The loop groups for the principal chiral model and the
sine-Gordon model are unitary on the real axis. Because of this unitarity 
condition a general meromorphic loop can be expressed as a product of
factors of a  particularly
simple form. However for the affine Toda theories the Toda fields are no 
longer real, and hence the unitarity condition on the real axis must be 
abandoned and we must look at a different class of meromorphic loops.
In this case we no longer have the factorisation theorem to allow the 
factorisation of general meromorphic loops into simple factors. However
in this paper we study a smaller 
class of meromorphic loops for which a factorisation 
result holds. Even though this class is relatively simple, we still find
new single soliton solutions to the affine Toda theories, as compared with
previous methods. These exhibit internal degrees of freedom which were not 
present in the previous solutions. However by setting these internal degrees 
of freedom to zero we recover the previously known solutions, and
their topological charges are also the same.

Affine Kac-Moody algebras are formed from loop algebras by inserting a central
extension. Using this central extension we give a very simple formula
in section 10 for the integrated energy-momentum density (and hence
the mass) 
of a soliton in terms of the central part  of a combination of meromorphic
loops.\vskip 0.2in

\noindent Contents:\\ \\ 
1. Introduction \\
2. The affine Toda field theories\\
3. The linear system and the symmetry condition\\
4. Group doublecrossproducts and integrable systems\\
5. The space-time evolution of $\phi(\lambda)$\\
6. Complex analysis of the loop group $\widehat{GL(n,\Bbb C)}$\\
7. Rank one projections and the soliton solutions\\
8. Multi-soliton solutions\\
9. Mass and topological charge of the single soliton solutions\\
10. Energy-momentum and the central extension\\
11. Discussion and conclusions\\

\resection{The affine Toda field theories}
The affine Toda field theories have equations of motion based on the
root system of an affine algebra $\hat{g}$. Let $\alpha_i, i=1,\ldots,
r$ be the simple roots of the Lie algebra $g$, and $\alpha_0$ be minus
the highest root $-\psi$, then the equations of motion of the affine
Toda field theories, for $u(x,t)$ an $r$-component scalar field,
are \beq
{\partial^2 u\over\partial t^2}-{\partial^2 u\over\partial x^2}
+{4\mu^2\over\beta}\sum_{i=0}^r
m_i{\alpha_i\over\alpha_i^2}e^{\beta\alpha_i.u} =0.\label{eq: Toda}\eeq
Here $m_i$ are certain integers such that 
$${\psi\over\psi^2}=\sum_{i=1}^r m_i{\alpha_i\over\alpha_i^2},$$
and $m_0=1$, so that $\sum_{i=0}^r m_i{\alpha_i\over\alpha_i^2}=0$. If
$\beta$ is purely imaginary then we see that a constant $u$ such
that $\alpha_i\cdot u\in {2\pi\over |\beta|}{\Bbb Z}$ is a
solution. These constant values make up the weight lattice of the
coroot algebra $\Lambda_{W}(g^{\vee})$, $u\in {2\pi\over
|\beta|}\Lambda_{W}(g^{\vee})$, and these values lie at the degenerate
minima of the potential in the Lagrangian to the affine Toda
model. Hence we expect soliton solutions to exist interpolating
different values of these minima at $x\rightarrow \infty$ and
$x\rightarrow -\infty$. Some solutions were first found by Hollowood
\cite{H1} for the $A_{n-1}$ theories using the Hirota method.
This method however is somewhat unsatisfactory and a more
powerful method exploiting the representation theory of affine
Kac-Moody algebras, and the vertex operators, was developed for all
simply-laced algebras in
\cite{OTUa, OTUb} by Olive, Turok and Underwood. These methods agreed
on what the solutions were for
the $A_{n-1}$ theories, it was thought that the single soliton
solutions were, for the species $i=1,\ldots,r$ of soliton, associated
with a node on the Dynkin diagram, and $\lambda_j$ a fundamental
weight of $g$,
\beq
e^{-\beta\lambda_j\cdot u}={1+Q_1\omega^{ij}W_i\over
1+Q_1W_i}\label{eq: Ansol}, \eeq
where \beq\omega=e^{2\pi i\over n},\qquad
W_i(\theta)=e^{m_i(e^{-\theta}x_+-e^{\theta}x_-)},\qquad x_\pm=t\pm x
\label{eq: W}\eeq
$$m_i=2\mu\sin\bigl({\pi i\over n}\bigr),\qquad Q_1\in {\Bbb C},$$
and $\theta$ is the rapidity of the soliton. The position of the
soliton is proportional to $\log|Q_1|$. The
solution (\ref{eq: Ansol}) is also singular at particular values of
the phase of $Q_1$ given by the zeroes of either the numerator or the
denominator of (\ref{eq: Ansol}). In this case the singular values of
$Q_1$ are simply given by straight lines joining the origin and
extending to infinity. For simplicity, we restrict our attention to
$n=4$, and $i=1$, then the singularities in the $Q_1$ plane are with
phases 
$0,\frac\pi 2,\pi,\frac{3\pi}2$.

The topological charge of the soliton is defined as
$$T={|\beta|\over 2\pi}\Bigl(u(\infty,t)-u(-\infty,t)\Bigr)\in
\Lambda_W(g^\vee),$$
and is independent of time $t$. The topological charge is a continuous
function of $Q_1$ and takes discrete values, so it will be constant as
we vary $Q_1$ within each of the four regions, but will not be
defined if $Q_1$ touches the singular lines. In fact we expect the
charge to jump and take a different value as $Q_1$ moves between the
regions  separated by the singularities. This is
corroborated in detail by work done by McGhee \cite{McGhee} for the $A_n$
theories, where the charges are computed. For the remaining Toda
theories, and the previously known single soliton solutions, the
singular regions in the $Q$ plane are also given by straight lines
joining the origin and extending to infinity, but the analysis is
slightly more complicated because more than one power in $QW$ is
present in the numerator and denominator of
$e^{-\beta\lambda_j\cdot u}$, compare with the formula (\ref{eq:
Ansol}) for $A_n$. We shall see that this will not be the case in
general for even the simpler $A_n$ theories. 

These methods \cite{H1,OTUa,OTUb} 
also agreed on a form for the two-soliton solution. Here we follow the
Olive-Turok-Underwood formalism
\cite{OTUa,OTUb}, where the two-soliton solution is understood in terms of a
special function $X^{jk}(\theta_j-\theta_k)$, which is a function
obtained when we normal order the two vertex operators
$F^{j}(\theta_j)$ and $F^{k}(\theta_k)$ associated with the solitons
of species $j$ and $k$ in the two-soliton solution. Here $\theta_j$
and $\theta_k$ are the rapidities of the two solitons which must be
real in order to make physical sense.
$$F^{j}(\theta_j)F^{k}(\theta_k)=X^{jk}(\theta_j-\theta_k):F^{j}(\theta_j)
F^{k}(\theta_k):$$
and $X^{jk}(\theta)$ can be given explicitly by 
\beq X^{jk}(\theta)=\prod_{p=1}^h\Bigl(1-e^\theta e^{{\pi i\over
h}(2p+{c(j)-c(k)\over 2})}\Bigr)^{\gamma_j\cdot\sigma^p\gamma_k}.\label{eq:
X} \eeq
Here $h$ is the Coxeter number of the Lie algebra, 
$c(j)=\pm 1$ is a particular `colour' depending on a
bi-colouration of the Dynkin diagram of $g$, where the soliton of
species $j$ is associated with a node of the Dynkin
diagram. Also $\gamma_j=c(j)\alpha_j$, and $\sigma$ is a special element of
the Weyl group known as the Coxeter element \cite{OTUa,OTUb}. For the
$A_{n-1}$ theories the two-soliton solution (species $j$ and species $k$)
is
\beq e^{-\beta\lambda_i\cdot u}={1+Q_1\omega^{ik}W_k+Q_2\omega^{ij}W_j+
X^{kj}(\theta_k-\theta_j)Q_1Q_2\omega^{i(k+j)}W_kW_j\over
1+Q_1W_k+Q_2W_j+
X^{kj}(\theta_k-\theta_j)Q_1Q_2W_kW_j}\label{eq: 2sol}\eeq
The coefficient $X^{kj}(\theta)$ tells us a surprising amount about
the interaction of two solitons, albeit where the single solitons are
the ones given by (\ref{eq: Ansol}) and not the more general ones
which we are about to discuss. The time delay experienced by the
soliton $k$ as it interacts with soliton $j$ is proportional to $\log
X^{kj}(\theta)$ \cite{FJKO}. $X^{kj}(\theta)$ can also be extrapolated
to the exact S-matrix of the solitons \cite{PRJ}.

In this paper we are not directly concerned with these properties of
$X^{kj}(\theta)$ related to the interaction of two
solitons, because we shall take the case where $X^{kj}(\theta)$
vanishes. This has not been treated in the literature before because 
it was previously thought that (\ref{eq: 2sol}) did not have
real total energy and momentum when $\theta_k-\theta_j$ is at the
zeroes of $X^{kj}(\theta)$. It was also thought in \cite{MIH} that the
restricted solution was singular. Some of these solutions were
mentioned by Caldi and Zhu \cite{ZC}, but not properly identified as 
true single solitons and also not fully discussed. 

\resection{The linear system and the symmetry condition}

The integrability of the affine Toda systems
follows from the zero-curvature condition \cite{OT1,OT2,OT3}
\beq
[\partial_++A_+,\partial_-+A_-]=0\label{eq: zcc},\eeq
where $A_{\pm}$ is given, in terms of the spectral parameter $\lambda$, by
\beq
A_{\pm}\ =\ \pm\frac12\beta\partial_\pm(u.H)\ \pm\ \lambda^{\pm 1}\mu
e^{\pm\frac12\beta u.H}E_{\pm 1}e^{\mp\frac12\beta u.H},
\label{eq: As} \eeq
and where $H$ is the Cartan-subalgebra of $g$, and
$$E_{+1}={E_{-1}}^\dagger=\sum_{i=0}^r\sqrt{m_i}E_{\alpha_i}\in g,$$ for
$E_\alpha$ the step operator in $g$ corresponding to the root
$\alpha$. From $\sum_{i=0}^r {m_i\alpha_i\over\alpha_i^2}=0$, 
it is easy to see that $[E_{+1},E_{-1}]=0$.
The compatibility condition
$\partial_+\partial_-\Phi=\partial_-\partial_+\Phi$ for $\Phi$ the
solution to the linear system
\beq \partial_\pm\Phi=\Phi A_\pm, \label{eq: lin_sys}\eeq
implies the zero-curvature condition (\ref{eq: zcc}), and here
$\Phi(x,t,\lambda)$ is valued in the loop group $\hat{G}$ of $\hat{g}$
\cite{PS}. Therefore any solution $\Phi$ to (\ref{eq: lin_sys}) 
in $\hat{G}$ will generate a solution to the affine Toda field
equations of motion (\ref{eq: Toda}). 

The simple vacuum solutions are given by $u$ constant with 
$e^{-\beta \lambda_i.u}=1$. In this case we define
\beq A_+=J(\lambda)=\mu\lambda E_{+1},\qquad
A_-=K(\lambda)=-\mu\lambda^{-1}E_{-1}\label{eq: JK_def}.\eeq
The equations 
$$\partial_+\Phi_0=\Phi_0J,\quad {\rm and}\quad \partial_-\Phi_0=\Phi_0K, $$
have the simple exponential solution 
$$\Phi_0=Ce^{Jx_+ + Kx_-},$$
if we subtract off this solution from $\Phi$ by setting
$$\phi=\Phi_0^{-1}\Phi,$$
we see that $\phi$ obeys the equations
\beq \partial_+\phi=\phi A_+ - J\phi,\qquad {\rm and}\quad
\partial_-\phi=\phi A_- - K\phi \label{eq: final_lin_system}. \eeq

We demand that $\lambda=0$ and $\lambda=\infty$ are regular points of
the function $\phi(\lambda)$. If we write (\ref{eq: final_lin_system}) as
$$\phi(\lambda)^{-1}\partial_+\phi(\lambda)
=A_+-\phi(\lambda)^{-1}J\phi(\lambda),\qquad
\phi(\lambda)^{-1}\partial_-\phi(\lambda)=
A_--\phi(\lambda)^{-1}K\phi(\lambda),$$
and set the singular parts of the right-hand sides of these
equations to zero as $\lambda\rightarrow 0$, we find that
\beq
\phi(0)^{-1}E_{-1}\phi(0)=e^{-\beta u.H/2}E_{-1}e^{\beta u.H/2},
\label{eq: regul0}\eeq
and for $\lambda\rightarrow\infty$, we find
\beq \phi(\infty)^{-1}E_{+1}\phi(\infty)=e^{\beta u.H/2}E_{+1}e^{-\beta u.H/2}.
\label{eq: regul00}\eeq

For the $A_{n-1}$ theories, in a basis where $H$ is diagonal, the
$n\times n$ matrices $E_{\pm 1}$ are defined as:
\beq
E_{+1}=\pmatrix{0 & 1 & 0 & \cdots & & 0 \cr & 0 & 1 & 0 & \cdots & \cr
& & 0 & 1 & 0 & \cdots \cr \vdots & & &\ddots & \ddots & \cr
0 & 0 & \cdots & &0& 1 \cr
1 & 0 & \cdots & & & 0 }\quad,\quad 
E_{-1}=E_{+1}^\dagger=\pmatrix{0 & 0 & \cdots &  & & 1\cr
1& 0 & \cdots & & & \cr
0& 1& 0 & \cdots & & \cr 0& 0& 1 &0 &\cdots & \cr
\vdots & & &\ddots & \ddots & \cr
0 & \cdots &  & & 1& 0 }.
\eeq
We also define the diagonal matrix
\beq
U=\pmatrix{1 & & & & \cr
             & \omega^{-1}& & & \cr
             & & \omega^{-2}& & \cr
             & & & \ddots & \cr
             & & & & \omega^{-(n-1)}},\eeq
where we recall that $\omega=e^{2\pi i\over n}$. With these
definitions for $U$ and $E_{\pm 1}$, it is easy to check the relations
\beq
UE_{\pm 1}U^\dagger=\omega^{\pm 1}E_{\pm 1}.\eeq
However these relations really follow since $U$ represents the action
of the Coxeter element $\sigma$ in the Weyl group, and $E_{\pm 1}$ are
elements of $g$ with principal grade $\pm 1$ respectively \cite{FLO,OTUa}.

If we conjugate the equations of motion (\ref{eq: final_lin_system}) 
by $U$, so that
$$U\partial_+\phi(\lambda) U^\dagger=U\phi(\lambda) A_+(\lambda)  
U^\dagger - U J(\lambda) \phi(\lambda)
U^\dagger,$$
since $[U,H]=0$ this implies
$$U\partial_+\phi(\lambda) U^\dagger= U\phi(\lambda) U^\dagger  
A_+(\omega\lambda) - J(\omega\lambda) U \phi(\lambda)U^\dagger,$$
and a similar equation for the $x_-$ component. The following  
important symmetry relation on $\phi(\lambda)$ is consistent with
these equations:
\beq U\phi(\lambda)U^\dagger =
f(\lambda)\phi(\omega\lambda),\label{eq: important2}\eeq
for some scalar $f(\lambda)$. We shall study the solutions
$\phi(\lambda)$ which satisfy a special case of this condition,
namely:
\beq U\phi(\lambda)U^\dagger\ =\ \phi(\omega\lambda),
\label{eq: important}\eeq

\resection{Group doublecrossproducts and integrable systems}
A doublecrossproduct of groups consists of two subgroups $\cal{M}$
and $\cal{G}$ of a larger group, which we call the doublecrossproduct
group ${\cal M}\bowtie{\cal G}$. The subgroups obey the restriction that
every element $x\in\cal{M}\bowtie\cal{G}$ can be uniquely factored as
$m\phi$, for some $m\in{\cal M}$ and $\phi\in{\cal G}$, and also can
be uniquely factored as $\psi n$, for some $n\in{\cal M}$, and
$\psi\in{\cal G}$. One of the simplest examples is the group of
permutations of the numbers $\{1,2,3\}$. We could take $\cal G$ to
consist of the identity and the permutation $(1,2)$, and $\cal M$ to
consist of the identity and both of the 3-cycles. This construction is
already familiar in the Hopf algebra literature, where it is used in
constructing non-commutative, non-cocommutative Hopf algebras
\cite{Shahn}. However we shall use it in the context of classical integrable
systems, where the Hopf algebra structure is not required.

Take a space-time $\cal S$, and a group doublecrossproduct 
${\cal M}\bowtie{\cal G}$. We are given a map $a:{\cal S}\rightarrow
{\cal M}$ called the classical vacuum map. The group $\cal G$ is
called the classical phase space. Then for any $s\in\cal S$, and some
initial $\phi_0\in\cal G$, we have the factorisation
\beq
a(s)\phi_0=\phi(s)b(s)\label{eq: group_fac}\eeq
where $\phi(s)\in\cal G$ gives the space-time evolution on the phase
space, and $b(s)\in\cal M$ contains all the information necessary to
reconstruct the classical solution.

Now assume that $\cal S$ is a manifold (alternatively we could have a
lattice, as in statistical mechanics) and that $\cal M$ and $\cal G$
are Lie groups. We can then differentiate the equation (\ref{eq: group_fac}) 
in a direction along space-time to obtain
\beq\partial_\xi\phi=(\partial_\xi a)a^{-1}\phi-\phi(\partial_\xi b)b^{-1},
\eeq
a form perhaps more instantly recognisable as the linear system for
inverse scattering (for example, equation (\ref{eq: final_lin_system})).

In inverse scattering typically the groups involved in the
doublecrossproduct are loop groups, and the equations for the
differentials then contain a complex parameter $\lambda$, the loop
parameter, this is also the $\lambda$ which appears in the explicit
linear system discussed in \S 3. To obtain solitons, $\cal M$ is taken
to be the group of functions from ${\Bbb C}^*$ to $GL_n({\Bbb C})$ which are
complex analytic (so that essential singularities are likely to be
present in elements of $\cal M$ at $0$ and $\infty$),
 and $\cal G$ is taken to be the group of meromorphic
functions from ${\Bbb C}_\infty$ to $GL_n({\Bbb C})$. In the principal
chiral model \cite{Edwin} or the sine-Gordon model \cite{Faddeev} the
group factorisation is exact, and a true group doublecrossproduct is
obtained. However the Toda case is much more interesting, since we do
not have the exact factorisation results to carry through all this
construction (this is due to the abandonment of unitarity in this
class of model). The factorisation can only be achieved in certain
special cases. We shall only present a single special case where the
factorisation definitely works, but which still gives some interesting
soliton  solutions to the model.

We shall show that the factorisation result applies for an analytic
function $a:{\Bbb C}^*\rightarrow GL_n({\Bbb C})$ and a meromorphic
loop of the form 
\beq\phi_0(\lambda)=\Biggl({\lambda\over\lambda-\alpha}P_0+
{\lambda\over\lambda-\omega\alpha}UP_0U^\dagger+\cdots +
{\lambda\over\lambda-\omega^{n-1}\alpha}U^{n-1}P_0{U^\dagger}^{n-1}\Biggr),
\label{eq: special_factor}\eeq
where $P_0$ is a rank one projection whose kernel contains all the
subspaces $U^{k}\,{\rm im}\,P_0$ for $1\leq k < n$. We would like to solve
the factorisation problem $a\phi_0=\phi b$ for a similar meromorphic
loop $\phi$ and a loop $b$ which is analytic on ${\Bbb C}^*$.

Let $P$ be the projection with the one-dimensional image 
$a(\alpha)\,{\rm im}\, P_0$, and kernel containing all the subspaces 
$U^k\,{\rm im}\,P$ for $1\leq k<n$. These conditions determine $P$ 
precisely, in the generic case where the subspaces $U^k\,{\rm im}\, P$,
$0\leq k < n$, span the whole space.
From this projection construct the
meromorphic loop
$$\phi(\lambda)=\Biggl({\lambda\over\lambda-\alpha}P+
{\lambda\over\lambda-\omega\alpha}UPU^\dagger+\cdots +
{\lambda\over\lambda-\omega^{n-1}\alpha}U^{n-1}P{U^\dagger}^{n-1}\Biggr).$$
Now we would like to show that the function $b=\phi^{-1}a\phi_0$ is
analytic on ${\Bbb C}^*$. The first step is to write down the inverse
as
$$\phi(\lambda)^{-1}=\Biggl({\lambda-\alpha\over\lambda}P+
{\lambda-\omega\alpha\over\lambda}UPU^\dagger+\cdots +
{\lambda-\omega^{n-1}\alpha\over\lambda}U^{n-1}P{U^\dagger}^{n-1}\Biggr),$$
which can be checked explicitly by calculation, and bearing in mind
that
$PU^kP=0$, for $1\leq k < n$, and $P+UPU^\dagger + \cdots + 
U^{n-1}P{U^\dagger}^{n-1}=1$. Now look at the part of 
$b=\phi^{-1}a\phi_0$ which might possibly be singular at
$\lambda=\alpha$,
$$
\Biggl({\lambda-\omega\alpha\over\lambda}UPU^\dagger +\cdots +
{\lambda-\omega^{n-1}\alpha\over\lambda}U^{n-1}P{U^\dagger}^{n-1}\Biggr)
a(\alpha){\lambda\over\lambda-\alpha}P_0.$$
This vanishes identically because of the condition on the kernel of
$P$, noting that ${U^\dagger}^k=U^{-k}=U^{n-k}$, for $1\leq k < n$,
so $b(\lambda)$ is analytic at $\lambda=\alpha$. Similar
arguments show that it is analytic at all $\omega^k\alpha$, so it is
analytic on ${\Bbb C}^*$. This proves the factorisation result.
\resection{The space-time evolution of $\phi(\lambda)$}
We will now prove that we can choose the classical vacuum map 
$$a(x,t)=e^{-J(\lambda)x_+-K(\lambda)x_-},$$
to solve the affine Toda field theories for the special case which we
have already discussed.

Now let $\phi(\lambda)=\phi_0(\lambda)\in{\cal G}$ be a product of
factors of the form (\ref{eq: special_factor}), 
at some initial value in space
and time. Perform the group factorisation order reversal operation
(\ref{eq: group_fac}) 
\beq e^{-Jx_+-Kx_-}.\phi_0(\lambda)=\phi(\lambda).b,\label{eq: ORO} \eeq
where $e^{-Jx_+-Kx_-}\in{\cal M}$, $\phi(\lambda)\in{\cal G}$ and
$b\in{\cal M}$. Given $e^{-Jx_+-Kx_-}$ and $\phi_0(\lambda)$ this
operation defines a unique $\phi(\lambda)$, which is a function of
space-time. Note that by definition $b$ is invertible. 

We can calculate $\phi(\lambda)$ from $\phi_0(\lambda)$ in the
following manner.
Suppose that the image of the residue of $\phi_0(\lambda)$ at a pole 
$\lambda=\alpha$ is $V_0\subset{\Bbb C}^n$, then from (\ref{eq: ORO})
the image of the
residue of $\phi(\lambda)$ at $\lambda=\alpha$ is
\beq V=e^{-J(\alpha)x_+-K(\alpha)x_-}V_0,\label{eq: evol}\eeq
since $b$ is invertible. However it should be noted that the
space-time evolution of the kernel of the residue is encoded in $b$,
and that this cannot be determined simply.

Now we shall prove that the $\phi(\lambda)$ satisfies the linear
system (\ref{eq: final_lin_system}).
We begin by differentiating (\ref{eq: ORO}) with respect to
$x_+$:
\bea 
-J\phi(\lambda)b&=&(\partial_+\phi(\lambda))b+\phi(\lambda)\partial_+b\cr\cr
-(\partial_+b)b^{-1}&=&\phi(\lambda)^{-1}J\phi(\lambda) + \phi(\lambda)^{-1}
\partial_+\phi(\lambda),\label{eq: b+}\eea
and similarly for $x_-$:
\beq -(\partial_-b)b^{-1}=
\phi(\lambda)^{-1}K\phi(\lambda)
+\phi(\lambda)^{-1}\partial_-\phi(\lambda)\label{eq: b-}.\eeq

If $\phi$ is a product of $m$ factors of the form 
(\ref{eq: special_factor}), then $\phi$ is regular and invertible at
$\lambda=\infty$, and near $\lambda=0$ it behaves as $\lambda^m\psi$,
where $\psi$ is regular and invertible at $\lambda=0$.

By examining the formula (\ref{eq: b+}) for $-(\partial_+b)b^{-1}$, we
see that it is regular at $\lambda=0$. Also from (\ref{eq: b+}), we
see that, by looking at the rate of growth of $-(\partial_+b)b^{-1}$ as
$\lambda\rightarrow\infty$,
\beq
-(\partial_+b)b^{-1}=D_+ + \lambda.C_+,\label{eq: rog} \eeq
where 
\beq C_+ = \phi^{-1}(\infty)\mu E_{+1}\phi(\infty) \label{eq: C+}\eeq
and 
\beq
D_+=\lim_{\lambda\rightarrow 0}\phi(\lambda)^{-1}\partial_+\phi(\lambda),
\label{eq: D+}\eeq
are independent of $\lambda$. Additionally since
$-(\partial_+b)b^{-1}$ satisfies the symmetry condition 
(\ref{eq: important}), we see that $D_+$ must be diagonal and $C_+$
only has non-zero entries where the matrix $E_{+1}$ is non-zero.

We repeat the analysis for $-(\partial_-b)b^{-1}$, and get 
\beq
-(\partial_-b)b^{-1}=D_-+\lambda^{-1}C_-\label{eq: b_-}.\eeq
where 
\beq D_-=\lim_{\lambda\rightarrow \infty}\phi(\lambda)^{-1}\partial_-
\phi(\lambda),
\label{eq: D-}\eeq
is diagonal and 
\beq C_- = -\lim_{\lambda\rightarrow 0}\phi^{-1}(\lambda)\mu
E_{-1}\phi(\lambda) \label{eq: C-}\eeq
only has non-zero entries where the matrix $E_{-1}$ is non-zero. 

For simplicity we will now consider a single pole case, with one
factor of the type (\ref{eq: special_factor}).
We separate out the diagonal terms of $\phi(\lambda)$ by setting $\phi
=\tilde{\phi}N$, where $N$ is diagonal and 
$$\tilde{\phi}(\lambda)=\sum_{k=0}^{n-1}{\lambda\over
\lambda-\omega^k\alpha}U^kaU^{-k},$$
with the diagonal elements of $a$ set equal to one. From  
(\ref{eq: explicit_sum}), $\phi(\infty)=N$, so 
\beq
C_+=N^{-1}\mu E_{+1}N \label{eq: C++},\eeq
and also from (\ref{eq: explicit_sum}) at $\lambda=0$ we see that

$$C_-\ = \qquad\qquad\qquad\qquad\qquad\qquad\qquad\qquad
\qquad\qquad\qquad\qquad\qquad
\qquad\qquad\qquad\qquad\qquad $$ 
$$-N^{-1}\pmatrix{0 & a_{12} & & & & \cr
                     & 0 & a_{23}& & &\cr
                     & &0 & a_{34}& &\cr
                     & & &\ddots & \ddots &\cr
                     & & &       &        & a_{n-1\,n}\cr
                  a_{n1}  & & & & & 0}^{-1}\mu E_{-1}
\pmatrix{0 & a_{12} & & & & \cr
                     & 0 & a_{23}& & &\cr
                     & &0 & a_{34}& &\cr
                     & & &\ddots & \ddots &\cr
                     & & &       &        & a_{n-1\,n}\cr
                  a_{n1}  & & & & & 0}N $$
\vskip 0.3in
$$=-N^{-1}\pmatrix{a_{n1} & & & \cr
                         & a_{12} & & \cr
                         &         & \ddots& \cr
                         &         & & a_{n-1\,n}}^{-1}\mu E_{-1}
\pmatrix{a_{n1} & & & \cr
                         & a_{12} & & \cr
                         &         & \ddots& \cr
                         &         & & a_{n-1\,n}}N. $$
From comparison with the linear system (\ref{eq: final_lin_system}),
we would like 
$$C_+=e^{\frac12\beta
u.H}\mu E_{+1}e^{-\frac12\beta u.H},\quad {\rm and}\quad
C_-=-e^{-\frac12\beta
u.H}\mu E_{-1}e^{\frac12\beta u.H}.$$
These will be satisfied if
$$e^{\beta u.H}=
\pmatrix{a_{n1} & & & \cr
                         & a_{12}  & &\cr
                          & & \ddots& \cr
                           & & & a_{n-1\,n}},$$
and $N=e^{-\frac12\beta u.H}$.

Now we shall check that this
is consistent with the constant diagonal
term in $A_\pm$, compare with equation (\ref{eq: As}):
\bea D_+&=&\lim_{\lambda\rightarrow 0}\phi^{-1}\partial_+\phi\cr\cr\cr
&=&\pmatrix{(\partial_+a_{n1})a_{n1}^{-1} & & & \cr
                         & (\partial_+a_{12})a_{12}^{-1} & &\cr
                         &         & \ddots& \cr
                         &         & & (\partial_+a_{n-1\,n})
a_{n-1\,n}^{-1}} + N^{-1}\partial_+ N \cr\cr\cr
&=&\beta(\partial_+u.H)-\frac12\beta(\partial_+u.H)\cr\cr
&=& \frac12\beta(\partial_+u.H).\eea
and similarly
\bea 
D_-&=&\lim_{\lambda\rightarrow \infty}\phi^{-1}\partial_-\phi\cr\cr
&=&N^{-1}\partial_- N \cr\cr
&=&-\frac12\beta(\partial_-u.H).\eea
This completes the proof that the $\phi(\lambda)$, found using the
order reversing operation (\ref{eq: ORO}), satisfies the equations of
the linear system (\ref{eq: final_lin_system}).

\resection{Complex analysis of the loop group $\widehat{GL(n,\Bbb C)}$}
Consider a meromorphic element  $\phi(\lambda)$ of the loop group 
$\widehat{GL(n,\Bbb C)}$ with a pole at $\lambda=\alpha_2$. The
inverse will have a pole at some $\lambda=\beta_2$. Hence we write 
\beq \phi(\lambda)=
\Bigl(A(\lambda)+{\lambda-\alpha_1\over\lambda-\alpha_2}a\Bigr)
\label{eq: A} \eeq
for some $\alpha_1\in\Bbb C$ and some matrix $a$, and the matrix
$A(\lambda)$ is regular at $\lambda=\alpha_2$. Provided the pole at 
$\lambda=\beta_2$ in the inverse of $\phi(\lambda)$ is simple, we can 
write,
\beq \psi(\lambda)=\phi(\lambda)^{-1}=
\Bigl(C(\lambda)+{\lambda-\beta_1\over\lambda-\beta_2}b\Bigr)
\label{eq: B} \eeq
for some matrix $b$ and with $C(\lambda)$ regular at
$\lambda=\beta_2$.

The conditions that $\phi(\lambda)\psi(\lambda)$ is regular at
$\lambda=\beta_2$ and at $\lambda=\alpha_2$ are respectively:
\beq
A(\beta_2)b+{\beta_2-\alpha_1\over\beta_2-\alpha_2}ab=0\label{eq:
one}\eeq
\beq
aC(\alpha_2)+{\alpha_2-\beta_1\over\alpha_2-\beta_2}ab=0,\label{eq:
twob}\eeq
eliminating $ab$ from (\ref{eq: one}) and (\ref{eq: twob}) gives
$${A(\beta_2)b\over\beta_2-\alpha_1}=-{aC(\alpha_2)\over\alpha_2-\beta_1},$$
implying  (provided $A(\beta_2)/(\beta_2-\alpha_1)$ is invertible)
\beq
b=-\Bigl({\beta_2-\alpha_1\over\alpha_2-\beta_1}\Bigr)A(\beta_2)^{-1}
aC(\alpha_2)
\label{eq: three}\eeq
substitution into (\ref{eq: twob}) implies
$$aC(\alpha_2)-{\beta_2-\alpha_1\over\alpha_2-\beta_2}aA(\beta_2)^{-1}a
C(\alpha_2)=0$$
which implies
$${\beta_2-\alpha_1\over\alpha_2-\beta_2}aA(\beta_2)^{-1}=\Bigl({\beta_2-
\alpha_1\over\alpha_2-\beta_2}aA(\beta_2)^{-1}\Bigr)^2,$$
so $$P={\beta_2-\alpha_1\over\alpha_2-\beta_2}aA(\beta_2)^{-1}$$ is a
projection, that is $P^2=P$.

This means that for 
\beq
a={\alpha_2-\beta_2\over\beta_2-\alpha_1}PA(\beta_2)\label{eq: four}
\eeq
to be non-trivial, $a$ must have rank less than $n$.
Since otherwise (if $a$ is invertible) then $P=1$ and from 
(\ref{eq: A}), $\phi(\beta_2)=0$. We shall see below that this
trivialises the solution.

We now impose the symmetry condition (\ref{eq: important})
$U\phi(\lambda)U^\dagger=\phi(\omega\lambda)$, which we assume to hold for
our solutions to the affine Toda linear system. The most general form for
$\phi(\lambda)$ with a pole at $\lambda=\alpha_2$, as above, is
\beq
\phi(\lambda)=\Bigl({\lambda-\alpha_1\over\lambda-\alpha_2}a+
{\lambda-\omega\alpha_1\over\lambda-\omega\alpha_2}UaU^\dagger+\cdots
+{\lambda-\omega^{n-1}\alpha_1\over\lambda-\omega^{n-1}\alpha_2}U^{n-1}a
{U^\dagger}^{n-1} + d\Bigr),\label{eq: final_form}\eeq
here $d$ is a diagonal matrix, so that $UdU^\dagger=d$. Hence we can
identify the $A(\lambda)$ from (\ref{eq: A}) with
\beq
A(\lambda)=\Bigl({\lambda-\omega\alpha_1\over\lambda-\omega\alpha_2}UaU^\dagger
+{\lambda-\omega^2\alpha_1\over\lambda-\omega^2\alpha_2}U^2a{U^\dagger}^2
+ \cdots +
{\lambda-\omega^{n-1}\alpha_1\over\lambda-\omega^{n-1}\alpha_2}
U^{n-1}a
{U^\dagger}^{n-1} + d\Bigr).\label{eq: Alambda}\eeq
This gives us an equation for $A(\beta_2)$ in terms of $a$ and $d$,
which can be used in conjunction with equation (\ref{eq: four}).

For $\alpha_1=0$, 
it is possible to show that, for $z=\lambda/\alpha_2$,
$$\Bigl({\lambda\over\lambda-\alpha_2}a+
{\lambda\over\lambda-\omega\alpha_2}UaU^\dagger+\cdots
+{\lambda\over\lambda-\omega^{n-1}\alpha_2}U^{n-1}a{U^\dagger}^{n-1}\Bigr)=$$
\beq{nz\over z^n-1}\pmatrix{z^{n-1}a_{11} & a_{12} &
z a_{13} &\cdots & \cdots & z^{n-2}a_{1n} \cr
z^{n-2}a_{21} & z^{n-1}a_{22} & a_{23} & z a_{24} & \cdots &
z^{n-3}a_{2n} \cr
z^{n-3}a_{31} & z^{n-2}a_{32} & z^{n-1}a_{33} & a_{34} & \cdots &
z^{n-4}a_{3n}
\cr
\vdots & & & & & \vdots \cr
z a_{n-11}&z^2a_{n-12}& \cdots & \cdots &z^{n-1}a_{n-1 n-1} & a_{n-1n} \cr
a_{n1}&z a_{n2}& \cdots & \cdots & z^{n-2}a_{n n-1} & z^{n-1}a_{nn} 
}.\label{eq: explicit_sum} \eeq 
Hence, if $\phi(\beta_2)=0$, then $a$ must be diagonal. 
Then $UaU^\dagger=a$, and
$\phi(\lambda)=f(\lambda)a$, for some scalar $f(\lambda)$. In this
case we can renormalise $\phi(\lambda)$ to $\phi(\lambda)=a$, and this
is trivial. Thus we conclude that $a$ must have a rank strictly less
than $n$. The simplest case to study is when $a$ is of rank one.

\resection{Rank one projections and the soliton solutions}

Consider the most general rank one matrix $a$
\bea
a&=&\pmatrix{v_1 & 0& & & 0\cr
           v_2 & 0& & & 0\cr
           \vdots & \vdots &\ddots & & \vdots\cr
           v_n & 0& & & 0} 
\pmatrix{w_1 & w_2 & \cdots & w_n\cr
         0   & 0   & \cdots & 0 \cr
         \vdots & \vdots & \ddots & \vdots\cr 
         0   & 0   & \cdots & 0}\cr\cr\cr
&=&\pmatrix{v_1w_1& v_1w_2 &\cdots &v_1w_n\cr
            v_2w_1 &v_2w_2  &      &v_2w_n \cr
            \vdots &        & \ddots &\vdots \cr
            v_nw_1 &\cdots  &     &v_nw_n },
\label{eq: a_rank_one} \eea 
and the diagonal matrix $d$, where $k$ is a constant
$$d=k\pmatrix{v_1w_1 & & & \cr
                     & v_2w_2 & & \cr
                     &        & \ddots & \cr
                     &        &        & v_nw_n}.$$
We also set $\alpha_1=0$.

Then with $\phi(\lambda)$ defined by (\ref{eq: final_form})
$${\rm det}\ \phi(\lambda)=v_1w_1v_2w_2\ldots v_nw_n
{\lambda^n-\beta_2^n\over \lambda^n-\alpha_2^n},$$
with $$\beta_2={\alpha_2k\over n+k}.$$
From (\ref{eq: Alambda}), we compute $A(\beta_2)$ and calculate 
$$P=aA(\beta_2)^{-1}{\beta_2\over \alpha_2-\beta_2},$$
we find that 
\beq
P=\frac1n\pmatrix{1 & \frac{v_1}{v_2} & \frac{v_1}{v_3} & \cdots & 
\frac{v_1}{v_n}\cr
    \frac{v_2}{v_1} & 1 & \frac{v_2}{v_3} & \cdots & \frac{v_2}{v_n}\cr
\frac{v_3}{v_1} & \cdots & & & \frac{v_3}{v_n}\cr
\vdots          &        & & \ddots & \vdots \cr
\frac{v_n}{v_1} & \frac{v_n}{v_2} & \cdots & & 1}.\label{eq: P}\eeq
This is a very special projection where the kernel is completely
fixed in terms of the image. Note that the kernel of $a$ is arbitrary,
it is the $n-1$ dimensional space orthogonal to the vector
$(w_1,w_2,\ldots,w_n)^{\rm T}$, 
and that we have eliminated $w_i$ from $P$.
This will be particularly advantageous for us, because we have seen
that it  is difficult
to solve the space-time dependence of the kernel, but rather simple to
solve for the image.

Now  observe that with this $P$, defined by (\ref{eq: P}),
$PU^{r}P=0$, for $r=1,2,\ldots, n-1$. Then (\ref{eq: four}) and 
(\ref{eq: Alambda}) imply
\beq
a=P{\alpha_2-\beta_2\over\beta_2}\Bigl(
{\beta_2\over\beta_2-\omega\alpha_2}UaU^\dagger + \cdots 
+{\beta_2\over\beta_2-\omega^{n-1}\alpha_2}U^{n-1}a{U^\dagger}^{n-1}+d\Bigr),
\label{eq: rhs} \eeq
after noting that
$$a={\alpha_2-\beta_2\over\beta_2}PA(\beta_2)$$ on the
right-hand side of this equation (\ref{eq: rhs}), we have
$$a={\alpha_2-\beta_2\over\beta_2}Pd = \frac{n}{k}Pd. \label{eq: five}$$
This last condition, $a=\frac{n}{k}Pd$, can be checked explicitly by
multiplying out the matrices. This is extremely useful for us, since
$UdU^\dagger=d$, and then
$$\phi(\lambda)=n\Bigl({\lambda\over\lambda-\alpha_2}P+
{\lambda\over\lambda-\omega\alpha_2}UPU^\dagger+\cdots
+{\lambda\over\lambda-\omega^{n-1}\alpha_2}U^{n-1}P{U^\dagger}^{n-1}
+ \frac{k}{n}1\Bigr)d',$$
where $$d'=\frac{d}{k}=\pmatrix{v_1w_1 & & & \cr
                     & v_2w_2 & & \cr
                     &        & \ddots & \cr
                     &        &        & v_nw_n}.$$
Hence this move effectively allows us to eliminate the kernel of $a$
from the discussion.
In this paper we shall restrict ourselves to the special case of the
limit $k\rightarrow 0$, to eliminate the diagonal term. Then
\beq
\phi(\lambda)=n\lambda\Bigl({1\over\lambda-\alpha_2}P+
{1\over\lambda-\omega\alpha_2}UPU^\dagger+\cdots
+{1\over\lambda-\omega^{n-1}\alpha_2}U^{n-1}P{U^\dagger}^{n-1}\Bigr)d',
\label{eq: final_phi}\eeq
where $P$ is still of the form (\ref{eq: P}).

This $\phi(\lambda)$, (\ref{eq: final_phi}), has the explicit inverse
\beq
\phi(\lambda)^{-1}={1\over n\lambda}d'^{-1}\Bigl((\lambda-\alpha_2)P+
(\lambda-\omega\alpha_2)UPU^\dagger+\cdots
+(\lambda-\omega^{n-1}\alpha_2)U^{n-1}P{U^\dagger}^{n-1}\Bigr).
\label{eq: inverse}\eeq

We are now in a position where we can study the soliton solutions
generated by a single simple pole factor (\ref{eq: final_phi}). These
will be single solitons. We adjust the normalization on the right-hand
side of (\ref{eq: final_phi}) so that
\beq\phi(\lambda)=\lambda\Bigl({1\over\lambda-\alpha_2}P+
{1\over\lambda-\omega\alpha_2}UPU^\dagger+\cdots
+{1\over\lambda-\omega^{n-1}\alpha_2}U^{n-1}P{U^\dagger}^{n-1}\Bigr)e^{-\beta
u.H/2}.\label{eq: final_final} \eeq
Then $\phi(\infty)=e^{-\beta u.H/2}$ and the regularity condition
(\ref{eq: regul00}) at $\lambda=\infty$ is automatically
satisfied. Writing $\phi(\lambda)=\tilde{\phi}(\lambda)e^{-\beta
u.H/2}$, where
$$\tilde{\phi}(\lambda)=\lambda\Bigl({1\over\lambda-\alpha_2}P+
{1\over\lambda-\omega\alpha_2}UPU^\dagger+\cdots
+{1\over\lambda-\omega^{n-1}\alpha_2}U^{n-1}P{U^\dagger}^{n-1}\Bigr),$$
and substituting into the regularity condition at $\lambda=0$,
(\ref{eq: regul0}), we find
\beq
\tilde{\phi}(0)^{-1}E_{-1}\tilde{\phi}(0)=e^{-\beta u.H}E_{-1}e^{\beta
u.H}\label{eq: modified_regul0}\eeq

We observed in section 5 that the space-time evolution of the image of
$P$, 
$${\rm im}\ P=<(v_1,v_2,\ldots,v_n)^{\rm T}>,$$ is very simple. It has the
evolution given by equation (\ref{eq: evol}). Given this image,
equation (\ref{eq: modified_regul0}) implies that
\beq
e^{-\beta \lambda_i.u}={v_{i+1}\over v_1}\label{eq: single_sol},\eeq
where $\lambda_i$ is the $i^{\rm th}$ fundamental weight. This can be
established by applying the standard basis vectors $f_i=\delta_{ij}$
to both sides of (\ref{eq: modified_regul0}). On the right-hand side
we obtain $e^{\beta\alpha.u}$, where $\alpha$ is a root of the basic
representation of $su(n)$. We note that all the simple roots
$\alpha_j$ are contained in the roots of the basic representation,
and then use the Cartan matrix $K_{ij}$ of $su(n)$ to
write $e^{-\beta\lambda_i.u}$ in terms of $e^{\beta\alpha_j.u}$, where
the fundamental weights, $\lambda_i$, are written in terms of the simple
roots, $\alpha_j$, as $\lambda_i=(K^{-1})_{ij}\alpha_j$.

Let $e_i\,:i=0,\ldots,n-1$ be the simultaneous eigenvectors of $E_{+1}$
and $E_{-1}$ (recall that $[E_{+1},E_{-1}]=0$)
$$e_i=\pmatrix{ 1 \cr \omega^i \cr \omega^{2i}\cr \vdots \cr
\omega^{(n-1)i}},\qquad {\rm where}\quad E_{\pm 1}e_i=\omega^{\pm
i}e_i\,:i=0,\ldots,n-1.$$
Then 
$$e^{-\alpha\mu E_{+1}x_++\alpha^{-1}\mu E_{-1}x_-}e_i=
e^{-\mu\alpha\omega^ix_++\mu\alpha^{-1}\omega^{-i}x_-}e_i.$$
Let $$W=e^{-\mu\alpha(\omega^i-1)x_++\mu\alpha^{-1}(\omega^{-i}-1)x_-}.$$

The initial subspace 
\beq V_0=<e_0+Qe_i>\label{eq: not_general},\eeq
has the space-time evolution
\bea
V&=&e^{-\mu\alpha E_{+1}x_++\mu\alpha^{-1}E_{-1}x_-}V_0 \cr\cr
&=&<\pmatrix{1 \cr {1+\omega^{i}QW\over 1+ QW} \cr
{1+\omega^{2i}QW\over 1+ QW}\cr \vdots \cr {1+\omega^{(n-1)i}QW\over 1+ QW}}>,
\label{eq: 7.10}
\eea
so we derive the solution
$$e^{-\beta \lambda_j.u}={1+\omega^{ij}QW\over 1+ QW}.$$
This is the standard single soliton solution (\ref{eq: Ansol}) 
for a soliton of species
$i$ found using the alternative methods\cite{H1,OTUa,OTUb}.

We are allowed initial subspaces more general than (\ref{eq:
not_general}), for example 
$$V_0=<e_0+Q_ke_k+Q_je_j>,\quad k\neq j$$
is perfectly valid, and we find the solution
$$e^{-\beta \lambda_r.u}={1+\omega^{kr}Q_kW_k+\omega^{jr}Q_jW_j\over 
1+ Q_kW_k+Q_jW_j}.$$
$$W_k=e^{m_k(e^{-(\theta-{(j-k)\pi i\over n})}x_+-e^{(\theta-{(j-k)\pi
i\over n})}x_-)}
\qquad W_j=e^{m_j(e^{-\theta}x_+-e^{\theta}x_-)},$$
where the phase of the position of the pole at $\lambda=\alpha_2$, is
fixed by $\alpha_2=i\omega^{-j/2}e^{-\theta}$, 
to give a real total energy and momentum. Here $\theta$ is the real
rapidity.

It was not realised beforehand in \cite{H1,OTUa,OTUb} 
that these types of solution were
genuine single soliton solutions, these types of solution and their
importance will be discussed in section 9.
\resection{Multi-soliton solutions}
Multi-soliton solutions can be computed by multiplying the meromorphic
loops (\ref{eq: final_final}) together in the loop group, each factor 
introducing an additional soliton into the system. We shall do this
for the special case of $g=su(3)$, and for a product of two loops.
We write the product of two loops in the two different orderings 
\begin{eqnarray*}
\phi(\lambda)=\Bigl({\lambda\over\lambda-\alpha_1}P_1&+&
{\lambda\over\lambda-\omega\alpha_1}UP_1U^\dagger+\cdots
+{\lambda\over\lambda-\omega^{n-1}\alpha_1}U^{n-1}P_1{U^\dagger}^{n-1}\Bigr)\cr
&&\times\Bigl({\lambda\over\lambda-\alpha_2}P_2+
{\lambda\over\lambda-\omega\alpha_2}UP_2U^\dagger+\cdots
+{\lambda\over\lambda-\omega^{n-1}\alpha_2}U^{n-1}P_2{U^\dagger}^{n-1}\Bigr)
e^{-\beta u.H/2}\cr\end{eqnarray*}
\begin{eqnarray*}
=\Bigl({\lambda\over\lambda-\alpha_2}P_3&+&
{\lambda\over\lambda-\omega\alpha_2}UP_3U^\dagger+\cdots
+{\lambda\over\lambda-\omega^{n-1}\alpha_2}U^{n-1}P_3{U^\dagger}^{n-1}\Bigr)\cr
&&\times\Bigl({\lambda\over\lambda-\alpha_1}P_4+
{\lambda\over\lambda-\omega\alpha_1}UP_4U^\dagger+\cdots
+{\lambda\over\lambda-\omega^{n-1}\alpha_1}U^{n-1}P_4{U^\dagger}^{n-1}\Bigr)
e^{-\beta u.H/2}\cr\end{eqnarray*}
For $su(3)$ we write this as 
$$\phi(\lambda)=\phi_1(\lambda).\Bigl({\lambda\over\lambda-\alpha_2}P_2+
{\lambda\over\lambda-\omega\alpha_2}UP_2U^\dagger
+{\lambda\over\lambda-\omega^{2}\alpha_2}U^{2}P_2{U^\dagger}^{2}\Bigr)
e^{-\beta u.H/2}=$$
$$\Bigl({\lambda\over\lambda-\alpha_2}P_3+
{\lambda\over\lambda-\omega\alpha_2}UP_3U^\dagger
+{\lambda\over\lambda-\omega^{2}\alpha_2}U^{2}P_3{U^\dagger}^{2}\Bigr)
.\phi_4(\lambda)e^{-\beta u.H/2}$$
The argument (\ref{eq: ORO}) for the space-time evolution of $P$ only 
works for a
meromorphic loop ordered in the left-most position. Hence we know the
space-time evolution of $P_1$ and $P_3$ and from this we
can recover the space-time evolution of $P_2$.
We compare the residues at $\lambda=\alpha_2$,
$\lambda=\omega\alpha_2$ and at $\lambda=\omega^2\alpha_2$ respectively,
\beq
\phi_1(\alpha_2)(P_2)=(P_3)\phi_4(\alpha_2)
\label{eq: 1A}\eeq
\beq
\phi_1(\omega\alpha_2)(UP_2U^\dagger)=(UP_3U^\dagger)\phi_4(\omega\alpha_2)
\label{eq: 2A}\eeq
\beq
\phi_1(\omega^2\alpha_2)(U^2P_2U^{\dagger 2})=(U^2P_3U^{\dagger 2})
\phi_4(\omega^2\alpha_2)
\label{eq: 3A}\eeq
since $\phi_4(\alpha_2)$ is
invertible the image from equation (\ref{eq: 1A}) of $P_2$
is $$V_2=\phi_1(\alpha_2)^{-1}V_3,$$
this implies that the image of $UP_2U^{\dagger }$ is 
$$V_2'=UV_2=U\phi_1(\alpha_2)^{-1}V_3.$$
Now
$$U\phi_1(\alpha_2)U^\dagger=\phi_1(\omega\alpha_2)$$
so $V_2'=\phi_1(\omega\alpha_2)^{-1}UV_3$, but this agrees with
equation (\ref{eq: 2A}), the image of $UP_2U^{\dagger }$ is 
$$\phi_1(\omega\alpha_2)^{-1}{\rm Im}(UP_3U^{\dagger})=\phi_1
(\omega\alpha_2)^{-1}UV_3,$$
therefore equations (\ref{eq: 1A}) and (\ref{eq: 2A}) are equivalent
since the projections are unique after we have imposed the
conditions
$$P_2UP_2=P_2U^2P_2=0.$$
We can also show that  (\ref{eq: 1A}), (\ref{eq: 2A}) and
(\ref{eq: 3A}) are all equivalent, using the same arguments.

We now consider the species 1 -- species 1 solution, in this case
$P_1$ and $P_3$ are the one-soliton solutions given by
$$V_1=\pmatrix{1 \cr {1+\omega W_1\over 1+W_1} \cr 
{1+\omega^2 W_1\over 1+W_1}}, \quad
V_3=\pmatrix{1 \cr {1+\omega W_2\over 1+W_2} \cr 
{1+\omega^2 W_2\over 1+W_2}} $$
where $W_i=Q_ie^{m_i(e^{-\theta_i}x_+-e^{\theta_i}x_-)}$. We
compute\footnote{
Note that we do not have to explicitly invert the matrix
$\phi_1(\alpha_2)$, we can use
the formula (\ref{eq: inverse}).}
$$V_2=\phi_1(\alpha_2)^{-1}V_3,$$
so that for 
$$P_2=\frac13\pmatrix{1 & A_2^{-1} & B_2^{-1} \cr A_2 & 1 & A_2/B_2 \cr B_2 &
B_2/A_2 & 1},$$
we have
$$A_2=\biggl({1+\omega^2W_1\over 1+W_1}\biggl)\biggr(
{1+({-\omega\alpha_1+\alpha_2\over\alpha_2-\alpha_1})W_1+
({-\alpha_1+\omega\alpha_2\over\alpha_2-\alpha_1})W_2+
({-\omega\alpha_1+\omega\alpha_2\over\alpha_2-\alpha_1})W_1W_2\over
1+({-\alpha_1+\omega^2\alpha_2\over\alpha_2-\alpha_1})W_1+
({-\omega^2\alpha_1+\alpha_2\over\alpha_2-\alpha_1})W_2+
({-\omega^2\alpha_1+\omega^2\alpha_2\over\alpha_2-\alpha_1})W_1W_2}\biggl)
$$
$$B_2=\biggl({1+\omega^2W_1\over 1+\omega W_1}\biggl)\biggr(
{1+({-\omega^2\alpha_1+\omega\alpha_2\over\alpha_2-\alpha_1})W_1+
({-\omega\alpha_1+\omega^2\alpha_2\over\alpha_2-\alpha_1})W_2+
({-\alpha_1+\alpha_2\over\alpha_2-\alpha_1})W_1W_2\over
1+({-\alpha_1+\omega^2\alpha_2\over\alpha_2-\alpha_1})W_1+
({-\omega^2\alpha_1+\alpha_2\over\alpha_2-\alpha_1})W_2+
({-\omega^2\alpha_1+\omega^2\alpha_2\over\alpha_2-\alpha_1})W_1W_2}\biggl)$$
we redefine 
$$Q_1\rightarrow {\alpha_2-\alpha_1\over\omega^2\alpha_2-\alpha_1}Q_1,
Q_2\rightarrow
{\alpha_2-\alpha_1\over\alpha_2-\omega^2\alpha_1}Q_2,$$ and compute
from equation (\ref{eq: regul0}) 
$$e^{-i\lambda_1.u}=
\biggl({1+\omega W_1+\omega W_2+\omega^2 W_1W_2 
{(\alpha_2-\alpha_1)^2\over(\omega^2\alpha_2-\alpha_1)
(\omega\alpha_2-\alpha_1)}\over
1+W_1+W_2+W_1W_2 
{(\alpha_2-\alpha_1)^2\over(\omega^2\alpha_2-\alpha_1)
(\omega\alpha_2-\alpha_1)}}\biggr)$$
$$e^{-i\lambda_2.u}=
\biggl({1+\omega^2 W_1+\omega^2 W_2+ \omega W_1W_2 
{(\alpha_2-\alpha_1)^2\over(\omega^2\alpha_2-\alpha_1)
(\omega\alpha_2-\alpha_1)}\over
1+W_1+W_2+ W_1W_2 
{(\alpha_2-\alpha_1)^2\over(\omega^2\alpha_2-\alpha_1)
(\omega\alpha_2-\alpha_1)}}\biggr)$$
we set $\alpha_i=i\omega^{-1/2}e^{-\theta_i}$ (appropriate for the one-soliton
solution of species 1), and compare with the Olive-Turok-Underwood 
formalism\cite{OTUa,OTUb}. We see
that we have rederived the correct normal ordering factor (\ref{eq: X})
$$X^{11}(\theta_1-\theta_2)={(e^{-\theta_1}-e^{-\theta_2})^2\over
(e^{-\theta_1}-\omega e^{-\theta_2})(e^{-\theta_1}-\omega^2 e^{-\theta_2})}$$
and the solution is 
$$e^{-i\lambda_1.u}=
\biggl({1+\omega W_1+\omega W_2+\omega^2 X^{11}W_1W_2 
\over
1+W_1+W_2+X^{11}W_1W_2 }
\biggr)$$
$$e^{-i\lambda_2.u}=
\biggl({1+\omega^2 W_1+\omega^2 W_2+ \omega X^{11} W_1W_2 
\over
1+W_1+W_2+X^{11} W_1W_2} 
\biggr).$$
We likewise compute the species 1 -- species 2 two-soliton solution.
This means that
$P_1$ and $P_3$ are the one-soliton solutions given by
$$V_1=\pmatrix{1 \cr {1+\omega W_1\over 1+W_1} \cr 
{1+\omega^2 W_1\over 1+W_1}}, \quad
V_3=\pmatrix{1 \cr {1+\omega^2 W_2\over 1+W_2} \cr 
{1+\omega W_2\over 1+W_2}} $$
We compute
$$V_2=\phi_1(\alpha_2)^{-1}V_3,$$
and 
$$A_2=\biggl({1+\omega^2W_1\over 1+W_1}\biggl)\biggr(
{1+({-\omega\alpha_1+\alpha_2\over\alpha_2-\alpha_1})W_1+
({-\alpha_1+\omega^2\alpha_2\over\alpha_2-\alpha_1})W_2+
({-\omega\alpha_1+\omega^2\alpha_2\over\alpha_2-\alpha_1})W_1W_2\over
1+({-\alpha_1+\omega^2\alpha_2\over\alpha_2-\alpha_1})W_1+
({-\omega\alpha_1+\alpha_2\over\alpha_2-\alpha_1})W_2+
({-\omega\alpha_1+\omega^2\alpha_2\over\alpha_2-\alpha_1})W_1W_2}\biggl)
$$
$$B_2=\biggl({1+\omega^2W_1\over 1+\omega W_1}\biggl)\biggr(
{1+({-\omega^2\alpha_1+\omega\alpha_2\over\alpha_2-\alpha_1})W_1+
({-\omega^2\alpha_1+\omega\alpha_2\over\alpha_2-\alpha_1})W_2+
({-\omega\alpha_1+\omega^2\alpha_2\over\alpha_2-\alpha_1})W_1W_2\over
1+({-\alpha_1+\omega^2\alpha_2\over\alpha_2-\alpha_1})W_1+
({-\omega\alpha_1+\alpha_2\over\alpha_2-\alpha_1})W_2+
({-\omega\alpha_1+\omega^2\alpha_2\over\alpha_2-\alpha_1})W_1W_2}\biggl).$$
We redefine 
$$Q_1\rightarrow {\alpha_2-\alpha_1\over\omega^2\alpha_2-\alpha_1}Q_1,
Q_2\rightarrow
{\alpha_2-\alpha_1\over\alpha_2-\omega\alpha_1}Q_2,$$ and then we compute
$$e^{-i\lambda_1.u}=
\biggl({1+\omega W_1+\omega^2 W_2+ W_1W_2 
{(\omega\alpha_2-\alpha_1)(\alpha_2-\alpha_1)
\over(\omega^2\alpha_2-\alpha_1)^2}
\over
1+W_1+W_2+W_1W_2{(\omega\alpha_2-\alpha_1)(\alpha_2-\alpha_1)\over 
(\omega^2\alpha_2-\alpha_1)^2}
}\biggr)$$
$$e^{-i\lambda_2.u}=
\biggl({1+\omega^2 W_1+\omega W_2+ W_1W_2 
{(\omega\alpha_2-\alpha_1)(\alpha_2-\alpha_1)
\over(\omega^2\alpha_2-\alpha_1)^2}
\over
1+W_1+W_2+W_1W_2{(\omega\alpha_2-\alpha_1)(\alpha_2-\alpha_1)\over 
(\omega^2\alpha_2-\alpha_1)^2}
}\biggr).$$
We have again rederived the correct normal ordering factor (\ref{eq: X})
$$X^{12}(\theta_1-\theta_2)={(\omega\alpha_2-\alpha_1)(\alpha_2-\alpha_1)
\over(\omega^2\alpha_2-\alpha_1)^2}.$$
\resection{Mass and topological charge of the single soliton solutions}
\noindent{\bf The soliton mass}

In \cite{OTUa} it is shown how the integrated energy-momentum density 
$$P^{\pm}(x'_+,x'_-)=\int_{-\infty}^{x'} dxE^{\pm}(x,t),$$
where $E^{\pm}(x,t)$ is the energy-momentum density, computed from an
`improved' energy-momentum tensor (derived essentially from the Lagrangian) is
particulary simple when evaluated on the soliton solutions.
Olive, Turok and Underwood derive the formula, equation (5.4) of
\cite{OTUa}
$$P^+(x')=\partial_-C\Bigl|_{x=-\infty}^{x'}$$
$$P^-(x')=-\partial_+C\Bigl|_{x=-\infty}^{x'},$$
where $C$ is given by 
$$C=-\frac2\beta\sum_{i=0}^r\Lambda_i.\Phi.$$
Here $\Phi$ is the field of the extended Toda theories with two extra 
fields due to the central extension ($\xi$) and the derivation
($\eta$), $\Phi=(u,\eta,\xi)$. To get the affine Toda theories, we set $\eta=0$. Also
\bea \Lambda_i&=&(\lambda_i,m_i,0)\cr\cr
\Lambda_0&=&(0,1,0).\eea
For the solution of the form of a ratio of Tau functions
$$e^{-\beta\lambda_i.u}=\frac{\tau_i}{\tau_0^{m_i}},$$
and for the $A_n$ theories, under discussion here, where $m_i=1$, this
formula for $C$ is 
$$C=\frac2{\beta^2}\sum_{i=0}^r\log\tau_i.$$
For illustrative purposes, we compute this integrated density
for the standard single solitons of the form
$$e^{-\beta\lambda_i.u}={1+Q\omega^{ij}W_j\over 1+QW_j},$$
in this case
$$C=\frac2{\beta^2}\sum_{i=0}^{n-1}\log(1+Q\omega^{ij}W_j),$$
and we compute
\bea
P^{+}(x)&=&-\frac2{\beta^2}\Bigl(\sum_{i=0}^{n-1}{Q\omega^{ij}W_j\over
1+Q\omega^{ij}W_j}\Bigr)m_je^\theta\cr\cr\cr
P^{-}(x)&=&-\frac2{\beta^2}\Bigl(\sum_{i=0}^{n-1}{Q\omega^{ij}W_j\over
1+Q\omega^{ij}W_j}\Bigr)m_je^{-\theta}.\label{eq: C} \eea
The total energy and momentum is
\bea
P^+&=&P^+(\infty)=-{2n\over\beta^2}m_je^\theta\cr\cr
P^-&=&P^-(\infty)=-{2n\over\beta^2}m_je^{-\theta}.\eea
This shows that the mass of the soliton\footnote{recall that $\beta$ was chosen to be purely imaginary, so that this
quantity is positive} is $-{2n\over\beta^2}m_j$.

\noindent{\bf The two parameter case} 

The solution given by the initial subspace 
\beq V_0=<v_0+Q_1v_k+Q_2v_j>\label{eq: second_choice}\eeq 
is 
\beq
e^{-\beta\lambda_i\cdot u}={1+Q_1\omega^{ik}U+Q_2\omega^{ij}W_j(\theta_j)
\over 1+Q_1U+Q_2W_j(\theta_j)}, \label{eq: new_sol} \eeq
where $W_j(\theta)$ is given by 
$$W_j(\theta)=e^{m_j(e^{-\theta}x_+-e^{\theta}x_-)},\quad m_j=2\mu\sin\bigl({\pi j\over n}\bigr),$$
and we have
chosen the phase of $\alpha$ so that $\alpha
=i\omega^{-j/2}e^{-\theta}$, for $\theta$ a real rapidity. Then we have
\bea
U&=&e^{\mu2\sin\bigl({k\pi\over n}\bigr)(e^{{\pi i(j-k)\over
n}}e^{-\theta}x_+-e^{-{\pi i(j-k)\over
n}}e^{\theta}x_-)}\cr \cr
&=&W_k\Bigl(\theta-{\pi i(j-k)\over
n}\Bigr) \label{eq: notatrest}. \eea
For simplicity we now put the soliton at rest by setting $\theta=0$, and
then
\bea W_j&=&e^{m_j (2x)} \cr
U&=&e^{m_k\bigl(2\cos\bigl({(j-k)\pi\over
n}\bigr)x+i2\sin\bigl({(j-k)\pi\over n}\bigr)t\bigr)} \label{eq: atrest}
\eea
We can then perform a Lorentz transformation on this to get back to a
moving soliton with $\theta\neq0$.

We choose $j$ and $k$ so that 
$$m_j>m_k\cos\Bigl({(j-k)\pi\over
n}\Bigr),$$
hence for large $x$, $W_j$ will dominate  $U$ and
$e^{-\beta\lambda_i\cdot\phi}$ will have the same limit as when
$Q_1=0$. It is clear that the intermediate term in the Tau functions
$\tau_i$, given by $Q_1\neq 0$ does not affect the limits as $x\rightarrow
\pm\infty$ of (\ref{eq: C}), and we conclude
that the mass of the solution (\ref{eq: new_sol}) is real and the same
as for the case $Q_1=0$. This mass is 
$M_j=-{2n\over\beta^2}m_j=-{4n\over\beta^2}\mu\sin\bigl({j\pi\over n}\bigr)$, see \cite{OTUa}. If we
re-insert the $\theta$ dependence into $W_j$ and $U$, then the total
energy and momentum $P^\pm$ in light-cone co-ordinates is 
$P^\pm=M_je^{\pm\theta}$. This shows that the phase of $\alpha$
has been chosen correctly so that for $\theta$ real, we get real total
energy and momentum, and that $\theta$ agrees with the standard
definition of rapidity.

If we now take the two-soliton solution (\ref{eq: 2sol})
\beq e^{-\beta\lambda_i\cdot u}={1+Q_1\omega^{ik}W_k(\theta_k)+Q_2\omega^{ij}W_j(\theta_j)+
X^{kj}(\theta_k-\theta_j)Q_1Q_2\omega^{i(k+j)}W_k(\theta_k)W_j(\theta_j)\over
1+Q_1W_k(\theta_k)+Q_2W_j(\theta_j)+
X^{kj}(\theta_k-\theta_j)Q_1Q_2W_k(\theta_k)W_j(\theta_j)}.
\label{eq: 2solb}\eeq
This equation makes physical sense for $\theta_k$ and $\theta_j$ real,
but will still satisfy the equations of motion (\ref{eq: Toda}) after
analytically continuing $\theta_k$ and $\theta_j$. In particular we
can set $\theta_k-\theta_j$ to be at a zero of
$X^{kj}(\theta_k-\theta_j)$. For the $A_{n-1}$ theories it is known
from the formula (\ref{eq: X}) that this zero is at 
\beq \theta_k-\theta_j=i{\pi (k-j)\over n}.\label{eq: pos_zero} \eeq
Hence with this restriction, and with $\theta=\theta_j$, we recover the
formula (\ref{eq: new_sol}) from (\ref{eq: notatrest}) and (\ref{eq: 2solb}).
The total energy and momentum $P^\pm$ of the generic two-soliton solution
(\ref{eq: 2solb}) is from \cite{OTUa}
\beq P^\pm=M_je^{\pm\theta_j}+M_ke^{\pm\theta_k}.\label{eq: two} \eeq
When evaluated at the analytically continued values (\ref{eq:
pos_zero}), this does not agree with our energy-momentum formula 
$P^\pm = M_je^{\pm\theta_j}$ derived for the
restricted solution (\ref{eq: new_sol}). This is somewhat surprising,
but a closer examination of the proof of the two-soliton result
(\ref{eq: two}) in \cite{OTUa}
requires $X^{kj}(\theta)\neq 0$, so that the dominant term for large
$x$ in both the
numerator and denominator of (\ref{eq: 2solb}) is $W_kW_j$.

Also note that, as remarked upon in \cite{FJKO},
 the case $j=k$ is empty, because $X^{jj}(0)=0$, and 
we recover the
standard one-soliton solution in the form
$$e^{-\beta\lambda_i\cdot u}={1+(Q_1+Q_2)\omega^{ij}W_j\over
1+(Q_1+Q_2)W_j}.$$

We now study the singularities of the solution (\ref{eq: new_sol}), in
the same way that we saw the singularities of the standard single
soliton solutions. For
illustrative purposes we restrict our attention to $n=4, j=1, k=3$.
With these values $m_j>m_k\cos\Bigl({(j-k)\pi\over n}\Bigr)$ (actually $j$
and $k$ are anti-solitons of each other, and $m_j=m_k$). Suppose that
$Q_2$ is given and that ${\rm Re}\ Q_2>0$, and ${\rm Im}\ Q_2<0$, say, 
for definiteness. We then sketch in the $Q_1$ plane the values where
the numerator and denominator of (\ref{eq: new_sol}) vanish, in the
rest frame of the soliton ($\theta=0$), as before. We also implicitly
absorb the time dependence $e^{im_k2\sin{(j-k)\pi\over n}t}$ from
(\ref{eq: atrest}) into $Q_1$.
%
%
\psbild{ht}{1}{10cm}{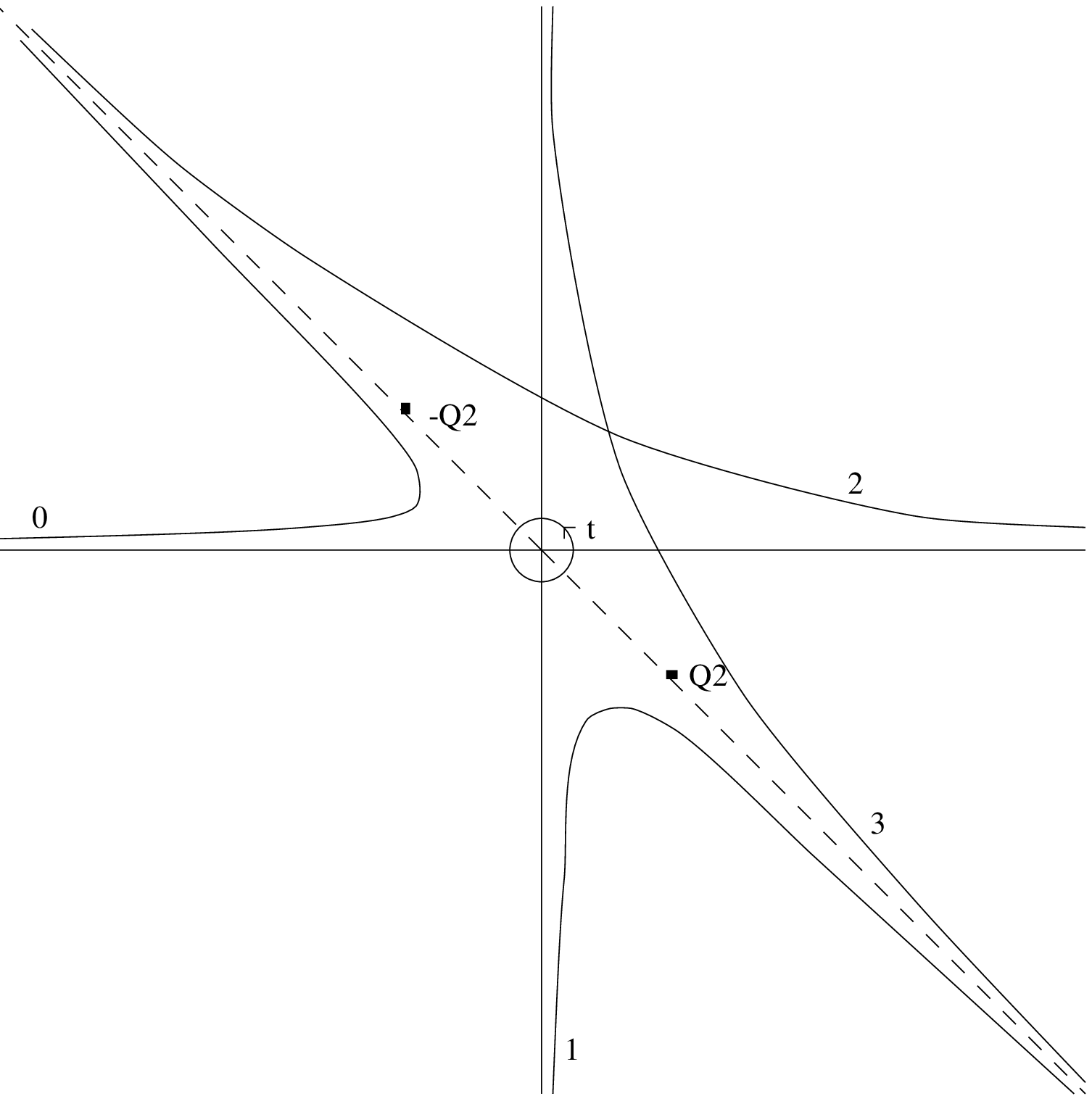}{Singularities
of (\ref{eq: new_sol}) with $n=4,j=1,k=3$ in the $Q_1$ plane}
\noindent The label $0$ refers to the zeroes of the 
denominator of (\ref{eq: new_sol}), and
the labels $j=1,2,3$ refer to the zeroes of the 
numerators of $e^{-\beta\lambda_j\cdot\phi}$.

 We see that provided $|Q_1|$ is sufficiently small, a circle of
radius $|Q_1|$ centered at the origin will not intersect any of the
singular curves, and hence the solution (\ref{eq: new_sol}) is free of
singularities for all time $t$. As $t$ increases (in the rest frame)
we move round the circle as shown. This time dependence is interesting
because it shows that in the rest frame of the soliton, it is not
completely motionless, in constrast to the case with $Q_1=0$,
and that there is an incipient small beating motion,
similar to the way that a breather breathes. This is shown in Figure
5. 
The breather is an
analytic continuation of a soliton--anti-soliton solution
(or of any two solitons with equal mass)
\cite{OTUb,MIH} which gives a real total energy and momentum, but
our solutions are certainly different from these breathers. Indeed in
\cite{MIH} it was thought that the analytic continuation of a two
soliton solution so that $X^{jk}(\theta)=0$ always gave a singular
solution, but the analysis in Figure 1 shows that this is not
necessarily the case.

Now Figure 1 was drawn given the assumption that $Q_2$ was known, and
that we had not accidently chosen a value which always gave a singular
solution. The phase of $Q_2$ is shown in the figure, and it is clear
by inspection that if we smoothly adjust the phase of $Q_2$ to either 
$0,\frac\pi 2,\pi,\frac{3\pi}2$, one of the four curves each in turn
will become a straight line passing through the origin, and the
solution must then be singular however small we choose $|Q_1|$. As the
phase of $Q_2$ becomes close to these four values we must choose
successively smaller values of $|Q_1|$ for the solution to be
non-singular. Hence in the $Q_2$ plane the singularities are the same
as the case with $Q_1=0$, but with the
understanding that $|Q_1|$ must be sufficiently small for a non-singular
solution. Also note that as $|Q_2|$ becomes smaller, but with the
phase of $Q_2$ fixed, the turning points of the curves in Figure 1 move
towards the origin, and tend to it in the limit when $|Q_2|\rightarrow
0$. Hence we must also adjust $|Q_1|$ depending on $|Q_2|$. 

We can also observe from Figure 1, that there are no new solutions for
any large values of $|Q_1|$, since the curves all go off to
infinity. A circle of large radius must intersect the curves.

Now a non-singular solution with appropriately small $|Q_1|$ is
continuously connected to the standard solution with $Q_1=0$,
therefore we
conclude from the continuity of the topological charge that the charge
is the same as the solution with $Q_1=0$. These are already well understood,
 at least for the $A_n$ theories, and have been calculated
\cite{McGhee}.

\noindent {\bf More than two parameters}

We pick a soliton species $j$, and consider all soliton species $k$
such that 
\beq m_j>m_k\cos\Bigl({(j-k)\pi\over n}\Bigr)\label{eq:
mass_condition}.\eeq
Define the set $B_j$ to consist of all the integers $k$ such that
(\ref{eq: mass_condition}) holds. The
number of possibilities for $k$ will depend on $j$, but as we have
seen in the first case, we can always take the anti-soliton $k$ to the
soliton $j$ provided $j\neq k$. If the anti-soliton species is the
same as the soliton species, then $n$ must be even (for the $A_{n-1}$
models), and
$j=\frac{n}2$. The property (\ref{eq: mass_condition}) must then be
true for all species $k$, since  the soliton $j$ is the heaviest. Also
note that for $j=1$ the only possibility for $k\in B_j$ is $k=n-1$.

We then choose the initial space
$$V_0=<v_0+\sum_{k\in B_j}Q_k v_k+Q_j v_j>$$
which generates the single soliton solution 
\beq
e^{-\beta\lambda_i\cdot u}={1+\sum_{k\in B_j}Q_k\omega^{ik}U_k+
Q_j\omega^{ij}W_j(\theta)\over
1+\sum_{k\in B_j}Q_kU_k+
Q_jW_j(\theta)}\label{eq: new_sol2}. \eeq
We have chosen the phase of $\alpha$ exactly as before, in order
to get a purely exponential behaviour of $W_j$ in $x$ and $t$, namely
$\alpha=i\omega^{-j/2}e^{-\theta}$, for $\theta$ real. Then 
\bea
U_k&=&e^{\mu2\sin\bigl({k\pi\over n}\bigr)(e^{{\pi i(j-k)\over
n}}e^{-\theta}x_+-e^{-{\pi i(j-k)\over
n}}e^{\theta}x_-)}\cr \cr
&=&W_k\Bigl(\theta-{\pi i(j-k)\over
n}\Bigr) \label{eq: notatrest2}, \eea
and $U_j=W_j$. The set $B_j$ has been defined so that $W_j$ dominates
for large $x$ over the purely exponential parts of $U_k$. This is
enough to guarantee that the mass of the solution (\ref{eq: new_sol2})
is $m_j$.

It is clear that we can also derive (\ref{eq: new_sol2}) from a
restriction of a multi-soliton solution. We take the multi-soliton
solution with solitons of species in $B_j$ and also of species $j$,
 and with separate rapidities
$\theta_k$, $k\in B_j$ and $\theta_j$.
 The multi-soliton solution is \cite{OTUb}
$$e^{-\beta\lambda_i\cdot u}={1+\sum_{k\in
B_j}Q_k\omega^{ik}W_k(\theta_k)+Q_j\omega^{ij}W_j(\theta_j)+\mbox{higher
terms}\over
1+\sum_{k\in B_j}Q_kW_k(\theta_k)+
Q_jW_j(\theta_j)+\mbox{higher terms}}.$$
The higher terms all involve more than one power of $W$. They are all
multiplied by products of $X$'s which vanish when we take 
\beq \theta_k-\theta_j=i{\pi(k-j)\over n}.\label{eq: dagger}\eeq
 For example, a term $W_{k_1}W_{k_2}$ for $k_1,k_2\in B_j$, is
multiplied by $X^{k_1 k_2}(\theta_{k_1}-\theta_{k_2})$, but from
(\ref{eq: dagger}) $\theta_{k_1}-\theta_{k_2}=i{\pi(k_1-k_2)\over
h}$, and $X^{k_1 k_2}(\theta_{k_1}-\theta_{k_2})$ has a zero at this
point. 

We can now discuss the structure of the singularities of (\ref{eq:
new_sol2}), after setting $\theta=0$. We pick an $r\in B_j$, and
assume that the remaining $Q_k, k\in B_j-\{r\}$ are given, but are very
small. We also assume that $Q_j$ is given and can be as large as we
please. As before, we absorb the time dependent phase of $U_r$ into
$Q_r$, but we cannot do this for the other phases.
We then sketch the singularities of (\ref{eq: new_sol2}). We first of
all sketch the curves for the situation $n=4, j=2, r=1$, with all
$Q_k=0$, and $k\in B_j-\{r\}$. This is done in Figure 2. Observe that the
singular phases of $Q_2$ are now $0$ and $\pi$. We now make 
$Q_k$, $k\in B_j-\{r\}$ non-zero, but very small. We fix time
$t$, and sketch the curves for all $x$. The curve will have the same
asymptotic behaviour as the previous case, but will have moved by a
small amount. In Figure 3, regions are  sketched approximately by
varying all $x$ and $t$. Hence the solution is singular at some 
$x$ and $t$ in the singular regions, but it may be possible for the
circle centered at the origin with the time dependent phase 
to avoid a singularity even though it may pass through a singular
region.  We shall analyse this situation below.
On the other hand, we certainly avoid
all singularities if $|Q_r|$
is sufficiently small so that $Q_r$ does not enter the singular
regions.  It is also clear that for the cases
where $Q_k, k\in B_j$ are small, the singular
lines in the $Q_j$ plane are the same as for the naive solution with 
$Q_k=0, k\in B_j$. 

\psbild{ht}{2}{10cm}{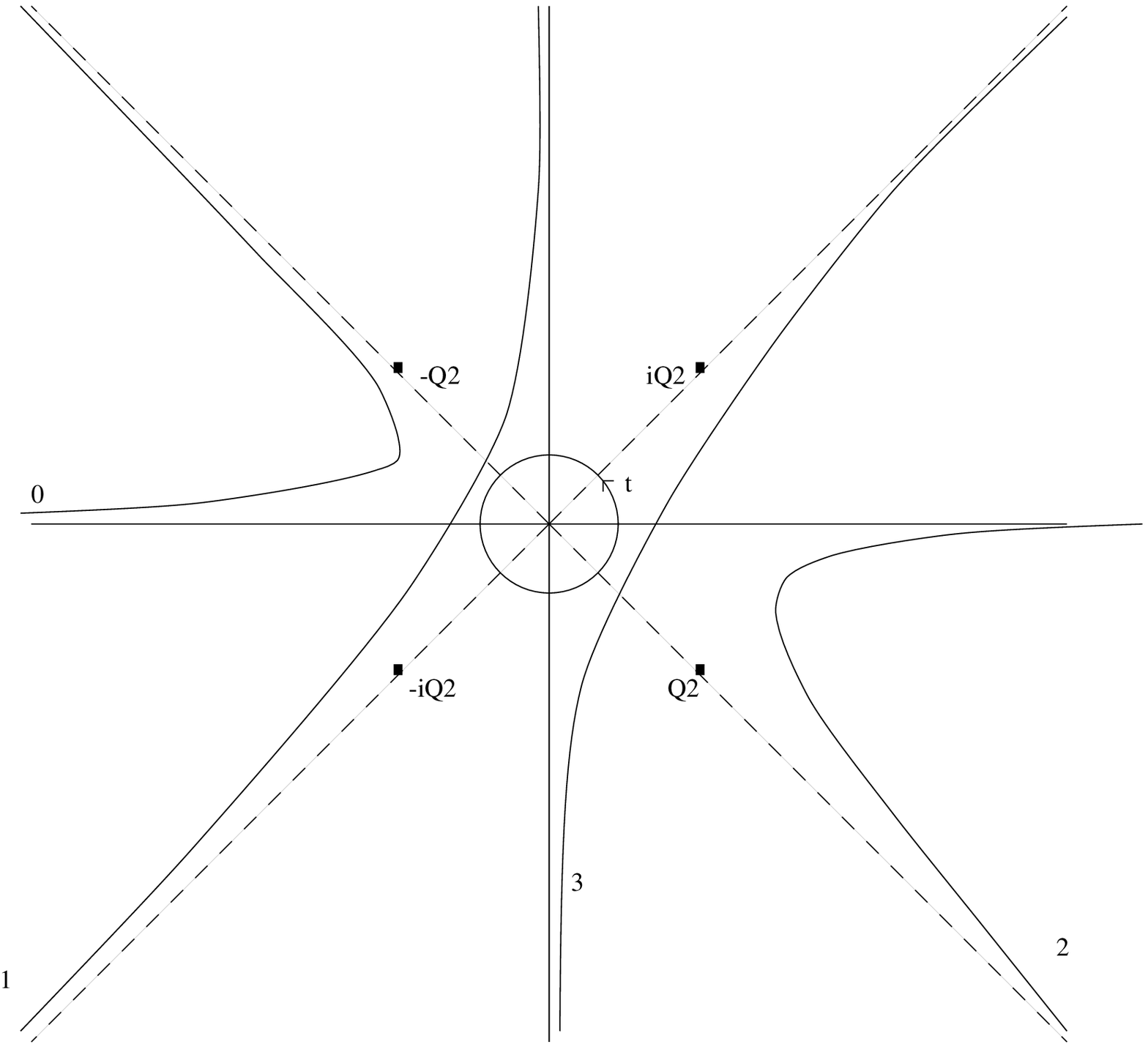}{Singularities
of $n=4, j=2, r=1$, in the $Q_1$ plane}
\psbild{ht}{3}{10cm}{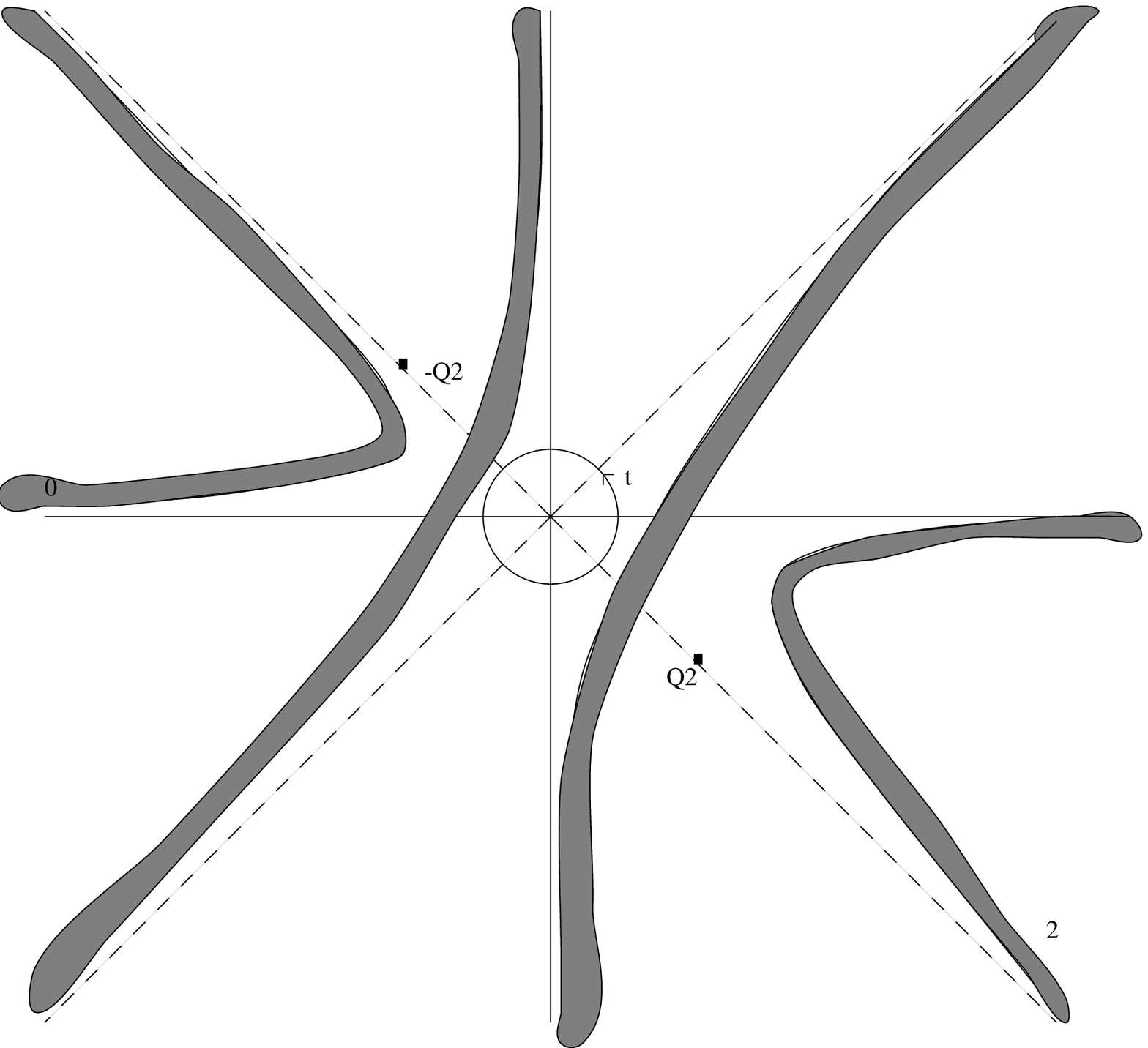}{Singularities
of $n=4, j=2, r=1, k=3$}

Now consider the case $n=4,j=2$ in more detail\footnote{This is the
simplest case where there is a possibility of finding new topological
charges.}, the solution is
\bea
e^{-\beta\lambda_1.u}={1+iQ_1U_1-iQ_3U_3-Q_2W_2\over
1+Q_1U_1+Q_3U_3+Q_2W_2}\cr\cr\cr
e^{-\beta\lambda_2.u}={1-Q_1U_1-Q_3U_3+Q_2W_2\over
1+Q_1U_1+Q_3U_3+Q_2W_2}\cr\cr\cr
e^{-\beta\lambda_3.u}={1-iQ_1U_1+iQ_3U_3-Q_2W_2\over
1+Q_1U_1+Q_3U_3+Q_2W_2}\label{eq: special_case}\eea
In this case $m_1=m_3$ (the solitons $1$ and $3$ are conjugate), and
at  $\theta=0$, 
\bea U_1&=&W_1(-\frac{\pi i}n)=
e^{m_1(e^{\pi i\over n}x_+-e^{-{\pi i\over
n}}x_-)}=e^{2m_1(\cos{\frac\pi n}x+i\sin\frac\pi n t)}\cr\cr
U_3&=&W_1({\pi i\over n})=e^{2m_1(\cos{\frac\pi n}x-i\sin\frac\pi n t)}.\eea
$U_1$ and $U_3$ have the same $x$ dependence, thus making it easier to
analyse the structure of the singularity. We analyse the zeroes of 
the numerator of the
first of the equations (\ref{eq: special_case}). These are given by
$$e^{-2m_1\cos{\frac\pi n}x}+\Bigl(iQ_1e^{(2m_1\sin\frac\pi
n)it}-iQ_3e^{-(2m_1\sin\frac\pi
n)it}\Bigr)-Q_2W_2e^{-2m_1\cos{\frac\pi nx}}=0,$$
or
\beq e^{-2m_1\cos{\frac\pi n}x}-Q_2W_2e^{-2m_1\cos{\frac\pi nx}}=
-\Bigl(iQ_1e^{(2m_1\sin\frac\pi
n)it}-iQ_3e^{-(2m_1\sin\frac\pi
n)it}\Bigr).\label{eq: compare} \eeq
We plot the left and right-hand sides of this equation (\ref{eq:
compare}) in Figure 4. Note there is no $x$ dependence in the
right-hand side, and no $t$ dependence in the left-hand side.
The closed curve given by the right-hand side must have a non-trivial
winding number around the origin, whatever complex values $Q_1$ and
$Q_3$ take, this curve is in fact an ellipse around the origin, which
gets bigger as we increase $Q_1$ and $Q_3$. We get similar pictures
for the remaining Tau functions of the solution (\ref{eq:
special_case}). However we cannot superimpose the pictures, as before,
since the ellipse is different for each Tau function. We conclude from
this that for large values of $Q_1$ and $Q_3$ we always get a singular
solution, and that the topological charge of the soliton cannot take
on any new values. We further conclude that an analysis of these
solutions for $n>4$ would probably also not find new topological charges,
although the analysis would be more complicated than the case presented
here.
\psbild{ht}{4}{10cm}{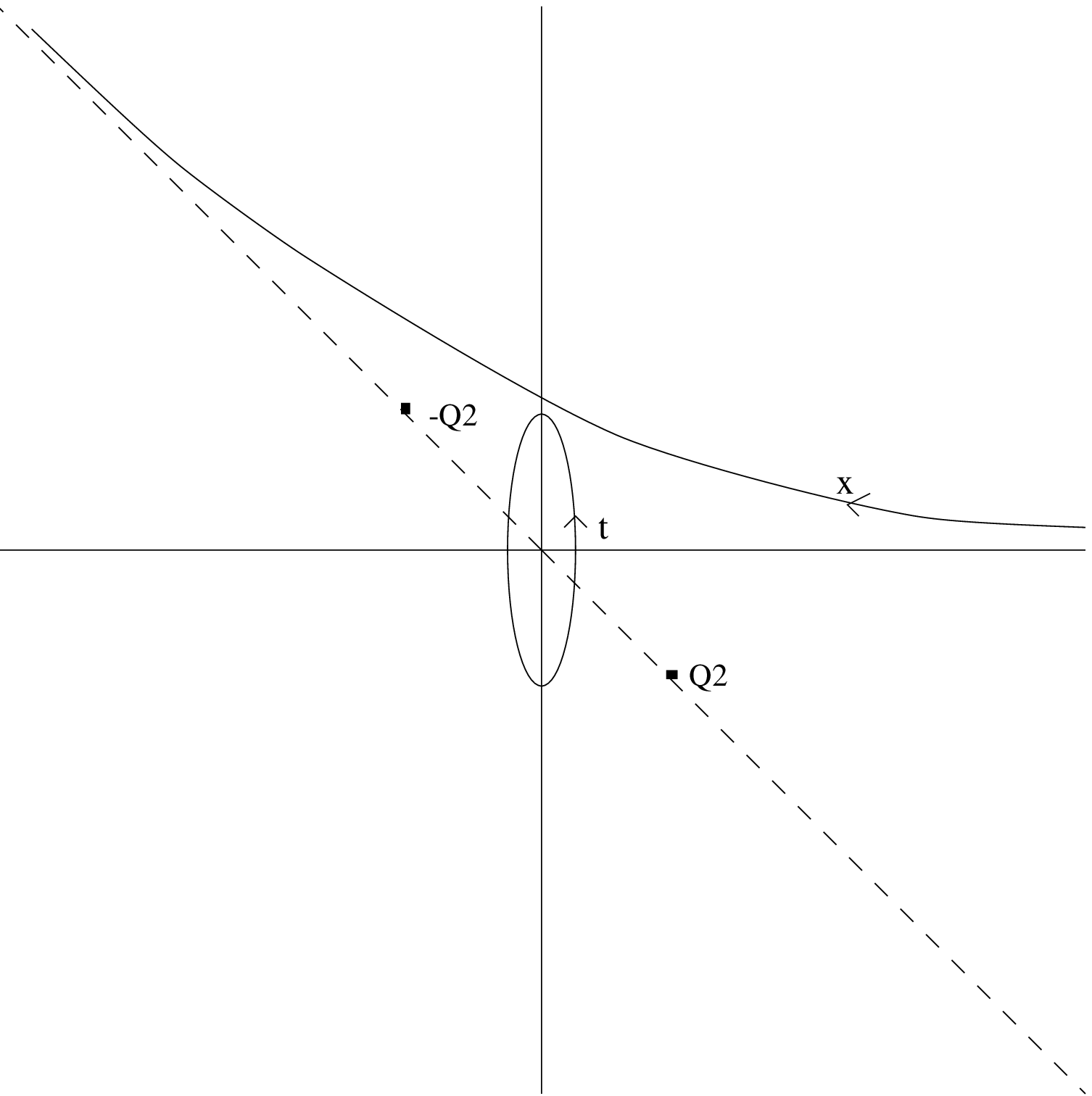}{Singularities of numerator of
$e^{-\beta\lambda_1.u}$, from (\ref{eq: special_case}) 
and (\ref{eq: compare})}

\psbild{ht}{5}{12cm}{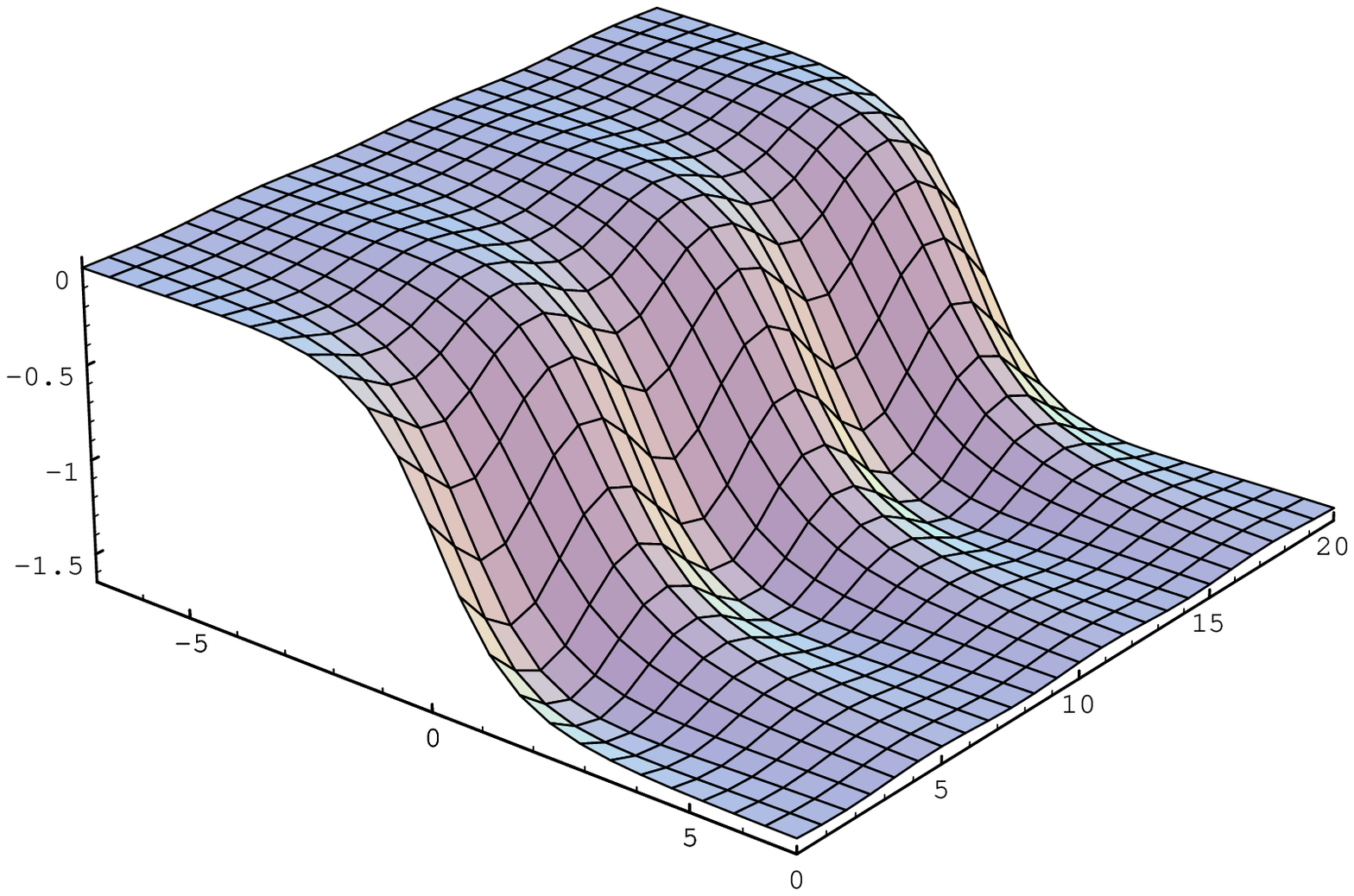}{Plot of ${\rm Re}(\lambda_1.\phi)$
(\ref{eq: new_sol})  for $n=3,j=1,k=3$}

\resection{Energy-momentum and the central extension}
Suppose that the Lie algebra $g$ has an adjoint invariant inner product
$\big<,\big>$, that is an inner product where
$$
\big<v,[u,w]\big>\ +\ \big<[u,v],w\big>\ =\ 0
$$
for all $u$, $v$ and $w$ in $g$. If there is a one parameter family of
 Lie algebra automorphisms $\theta_t$ ($t\in \Bbb R^+$) which preserve the
inner product (that is
$\big<\theta_t v,\theta_t w\big>=\big<v,w\big>$ for all $t\in \Bbb
R^+$, and $\theta_1 v=v$),
then we can define a Lie algebra 2-cocycle
$$
\omega(v,w)\ =\ \big<v,\frac{d}{dt}\Biggl|_{t=1} \theta_t w\big>\ .
$$
In our case we wish to look at cocycles on
the Lie algebra of the doublecrossproduct group
defined earlier, and in this case we can
define an inner product by the formula
$$\big<u,v\big>\ =\ \kappa\Im\Trace
\int_{\gamma}  \bar uv \frac {dz}z\ ,$$
where $\gamma$ is a contour in the complex plane and $\kappa$ is a
real constant which we shall assign a specific value to later.
The bar operation is defined by $\bar u(z)=Lu(\bar z)^*L^{-1}$, which makes
$\bar u$ into an analytic function. Here $L$ is the matrix
$$L=\pmatrix{0 & & & & 1\cr
               & & &1 & \cr
               & &1 & & \cr
               & \vdots & &  & \cr
              1 & & &  & 0 }.$$
This definition is designed so that $\bar{u}(z)$ is an analytic
function obeying the symmetry condition (\ref{eq: important}), if
$u(z)$ does. The key to showing this is the observation that
$L^{-1}UL=U^{-1}$. Then
\bea
Lu(\overline{\omega z})^*L^{-1}&=&Lu(\omega^{-1}\bar{z})^*L^{-1}\cr\cr
                          &=&LU^{-1}u(\bar{z})^*UL^{-1} \cr\cr
                          &=&ULu(\bar{z})^*L^{-1}U^{-1}. \eea

In our case we take $\gamma$ to consist of two circles, one anticlockwise of
large radius about the origin, and one clockwise of small radius
about the origin. The rather unusual looking imaginary part in the formula is
 accounted
for by the fact that the contour $\gamma$ is reversed under taking
its complex conjugate.
 The automorphism $\theta_t$ is defined by
$\theta_t(f)(z)=f(tz)$. If $t=e^{-\zeta}$, this is a Lorentz boost on
the inverse scattering system by a rapidity $\zeta$. 
To see that the inner product is invariant under
$\theta_t$, look at the integral
$$
\int_\gamma \theta_t(\bar uv)(z) \frac {dz}z\ =\
\int_{t\gamma} \bar uv(x) \frac {dx}x\ ,
$$
where we have made the substitution $x=tz$,
 and $t\gamma$ is just $t$ times
the old contour $\gamma$. If $u$ and $v$ are only
singular at $0$, $\infty$, and a finite number of poles in $\Bbb C^*$, then
we can deform the contour $t\gamma$ into $\gamma$ without crossing
any singularities, so the value of the integral is unaffected by $t$.
Now notice that the derivative $\frac d{dt}|_{t=1}\theta_t f(z)=z\frac d{dz}
f(z)$, so the cocycle becomes
$$
\omega(v,w)\ =\ \kappa\Im\Trace\int_\gamma \bar v(z) w'(z)\ dz.
$$
The reader may have noticed that the cocycle we have defined is
trivial over one of the subgroups ${\cal M}$ in the
doublecrossproduct. 
If we calculate
$\omega(v,w)$ for $v$ and $w$ analytic in $\Bbb C^*$, then we can continuously
deform the two circles in $\gamma$ together without meeting any
singularities, so their contributions to the integral cancel.

Now look at the central part of $\phi^{-1}J\phi$. We
turn to a formula in \cite{PS} at page 64 which tells us that this is given by
$$
\big<J\ ,\ \big(\frac{d}{dt}\Biggl|_{t=1} \theta_t \phi\big)\phi^{-1}
\big>\ .
$$
Now this can be rewritten in terms of the integral definition of the inner
 product, as
$$
\kappa\Im\Trace\int_\gamma
\phi'\phi^{-1}\overline{J}.dz\ .
$$
Note that any constant matrix multiplying $\phi$ on the right is
eliminated from this last equation.
This contour integral can be expressed as a sum of contributions from
 the residues at each pole of $\phi$. If we do this we get a formula analagous
to the Morse functions appearing in \cite{Edwin} and \cite{decay},
where they are
linked to the integrated energy and momentum density in the principal chiral
model.

Let us take the special case of an $n$-tuple of a single pole with
residue the rank one
projection, which gives some single soliton solutions.
As before,
$$\phi(z)\ =\ \frac{zP}{z-\alpha}\ +\ \frac{zUPU^{-1}}{z-
\omega\alpha}\ +\ \dots\ +\
\frac{zU^{n-1}PU^{1-n}}{z-\omega^{n-1}\alpha}\ ,
$$
and with a little calculation
$$
z\frac{d\phi}{dz}\phi^{-1}\ =\ -\ \sum_{k=0}^{n-1}
\frac{\omega^k\alpha}{z-\omega^k \alpha} U^k P U^{-k}\ ,
$$
so the central part of $\phi^{-1}J\phi$ is (using $J=\mu\lambda E_{+1}$)
$$-\ \kappa\mu\Im \sum_{k=0}^{n-1}\Trace\int_\gamma
\frac{\omega^k\alpha}{z-\omega^k \alpha} U^k P U^{-k}
\overline{z E_{+1}}.\frac{dz}{z}\qquad\qquad\qquad\qquad
\qquad\qquad\qquad\qquad$$
\bea
&=&\
-\ \kappa\mu\Im \sum_{k=0}^{n-1}\Trace\int_\gamma
\frac{\omega^k\alpha}{z-\omega^k \alpha} U^k P U^{-k}
LzE_{-1}L^{-1}.\frac{dz}{z} \cr \cr \cr
&=&\ -\ \kappa\mu\Im \sum_{k=0}^{n-1}\Trace\int_\gamma
\frac{\omega^{k}\alpha}{z-\omega^k \alpha} U^{k} P
U^{-k} E_{+1} U^{k} U^{-k}.dz \cr \cr \cr
&=&\ -\ \kappa\mu\Im \sum_{k=0}^{n-1}\Trace\int_\gamma
\frac{\alpha}{z-\omega^k \alpha}  P
E_{+1}.dz \cr \cr \cr
&=&\ -\ \kappa\mu\Im\  2\pi i\sum_{k=0}^{n-1}\Trace
{\alpha}  P
 E_{+1} \cr \cr \cr \ &=&\
-\  2\pi\kappa\mu n \ \Re\ {\alpha} \Trace
  P E_{+1} \ ,
\eea
and similarly the central part of $\phi^{-1}K\phi$ is
$$
-\  2\pi\kappa\mu n \ \Re\ \frac1{\alpha} \Trace
  P
 E_{-1} \ .
$$

We compute the central part of $\phi^{-1}K\phi$ for the single soliton
of type (\ref{eq: 7.10}). In the notation of (\ref{eq: P}), where
$P_{ij}=\frac1n\frac{v_i}{v_j}$:

$$\Trace(PE_{-1})=\frac1n\Bigl(\frac{v_1}{v_2}+\frac{v_2}{v_3}+\cdots 
+\frac{v_n}{v_1}\Bigr),$$
so the central part of $\phi^{-1}K\phi$ is (with $v_n=v_0$)
$$-2\pi\mu\Re\frac\kappa\alpha\Bigl(\sum_{i=0}^{n-1}\frac{v_i}{v_{i+1}}\Bigr),$$
setting $\alpha=i\omega^{-j/2}e^{-\theta},$
$$=2\pi\Re i\kappa\mu e^{\theta}\omega^{j/2}\Bigl(\sum_{i=0}^{n-1}{1+Q
\omega^{(i-1)j}W_j\over
1+Q\omega^{ij}W_j}\Bigr).$$
If we normalise this by subtracting off a constant so that it is zero
as $x\rightarrow -\infty$, then
\begin{eqnarray*}
&=&2\pi\Re i\kappa\mu
e^{\theta}\omega^{j/2}\Bigl(\sum_{i=0}^{n-1}{(1+Q\omega^{(i-1)j}W_j)
-(1+Q\omega^{ij}W_j)\over
1+Q\omega^{ij}W_j}\Bigr)\cr\cr\cr
&=&2\pi\Re i\kappa\mu
e^{\theta}\omega^{j/2}\Bigl(\sum_{i=0}^{n-1}{Q\omega^{ij}(\omega^{-j}-1)W_j\over
1+Q\omega^{ij}W_j}\Bigr)\cr\cr\cr
&=&2\pi\Re \kappa
e^{\theta}\Bigl(\sum_{i=0}^{n-1}{Q\omega^{ij}m_jW_j\over
1+Q\omega^{ij}W_j}\Bigr).\end{eqnarray*}
If we set $\kappa=-\frac{1}{\pi\beta^2}$, this agrees with the
integrated energy-momentum density, $P^+(x)$, formula (\ref{eq: C}),
provided we relax the real part condition. Observe that for $n>2$,
there is a non-zero imaginary part (which is zero in the limit
$x\rightarrow\infty$, so that the total mass is real). This imaginary
part occurs because these Toda models are not unitary. 

\resection{Discussion and conclusions}
One advantage of the inverse scattering method is that all single
solitons are generated by one uniform procedure. We have seen how the
new solutions with extra modes can be recovered from previous methods by 
restricting multi-soliton solutions. However it is not so obvious from these 
other methods that these solutions are indeed single solitons.

It is worth considering these new classical solutions in the context of the
quantum theory. Hollowood \cite{H2} has computed quantum corrections to the 
mass of the single soliton in these theories, using semi-classical
quantisation techniques.
As well as corrections from scattering states, with a continuous frequency,
(or from dispersive fluctuations),
Hollowood finds corrections from solitonic fluctuations around the
standard single soliton. He restricts a two-soliton solution in exactly the way
prescribed here, by setting $X^{ij}(\theta)=0$, to obtain a particular
solitonic fluctuation around the standard solution. This oscillates at
a certain discrete frequency $\omega$, and he suggests that the mass
receives a quantum correction of $\frac12 \hbar\omega$, the ground
state energy of the harmonic oscillator associated with the
fluctuation. We would 
arrive at the same answer as Hollowood for the mass correction
if we interpret  this as an internal oscillation  of the single soliton, 
rather than as a fluctuation away from it (but remaining within the
soliton  sector).
Semi-classically the higher states
of the harmonic oscillator correspond to excited soliton states with 
increasing masses higher than the ground state. These excited states are seen 
as a string of poles in the exact quantum S-matrix \cite{H3,PRJ} due to the 
fusing of solitons. These are present for all the $A_n$ theories when $n\ge 2$.
From the S-matrix there are a finite number of these excited solitons
associated with each species,
with masses dependent on the coupling in the model, their number
increasing to infinity as the coupling tends to 
zero. This semi-classical mass calculation \cite{H2} is followed up
for other affine Toda theories in \cite{WM}.

Types of meromorphic loop, more complicated than the simple rank one
case described here, could possibly
give new solutions, or find some of the missing topological charges. 
These missing
charges arise from the fact that for all the $A_n$ theories, we expect
that a soliton
species associated with a fundamental representation (or equivalently
a  node of the 
unextended Dynkin diagram of the Lie algebra) will fill all the
weights of  that 
representation. This is necessary for the quantum S-matrix formalism
to work \cite{PRJ}, and certainly necessary if 
there is a quantum group symmetry in the model, so that the solitons
transform in the fundamental representations under the action of the
symmetry.
At present it is only possible to show that a subset
of  the weights 
are filled\cite{McGhee} by classical single soliton solutions.
For large $n$ the number of missing charges
can be  considerable.

The reader should also compare our method with the inverse scattering
method for $A_n$ affine Toda theories as developed by
Niedermaier\cite{Nied}. In this paper Niedermaier shows through the
calculation of Jost functions and the Gelfand-Levitan-Marchenko
integral equations that the linear system in inverse scattering gives
an alternative derivation of the Tau functions as certain matrix
elements of a group orbit -- this plays a r\^{o}le in the
Olive-Turok-Underwood formalism.

However it remains to be seen to what extent the vertex operators in the
Olive-Turok-Underwood formalism are intimately related to
the meromorphic loops studied here.

In the inverse scattering method the central extension of the loop group
is not required to solve the classical equations of motion,
however if the central extension is introduced into the method, as
prescribed in this paper, an extremely simple formula for the
integrated energy-momentum density arises. This is interesting because
of the folklore that the energy of a representation is given by the
value of its central term (even in the classical case).
We expect that quantisation of the theory
would require a central extension on the loop group, because only then
is there a non-trivial positive-energy (highest weight) representation
theory.

The energy-momentum calculation shows that the cocycle giving the
central extension of the loop group depends on the coupling, $\beta$. This is
the only place where the coupling enters except for a global rescaling
of the Toda fields $u$, and we can expect that the coupling would
enter the quantum theory only through the action of the central
extension on the representation  of the loop group.

\vskip 0.4in
{\bf Acknowledgements}

One of us (PRJ) would like to thank Prof. David Olive for introducing
him to the affine Toda solitons. PRJ also acknowledges support from
EPSRC, in the form of a studentship under which a large part of this
work was carried out, and PPARC.

\end{document}